\documentclass[aps,prd,notitlepage,nofootinbib,showpacs,10pt,tightenlines]{revtex4-1}

\usepackage{amssymb,amsmath,bm}
\usepackage{hyperref} 

\usepackage{graphics, graphicx}
\usepackage{color}

\newcommand{\be}{\begin{equation}}
\newcommand{\ee}{\end{equation}}
\newcommand{\bea}{\begin{eqnarray}}
\newcommand{\eea}{\end{eqnarray}}
\newcommand{\bem}{\begin{multline}}
\newcommand{\eem}{\end{multline}}
\newcommand{\beg}{\begin{gather}}
\newcommand{\eeg}{\end{gather}}

\newcommand{\as}{\alpha_s}

\def\eq#1{{Eq.~(\ref{#1})}}
\def\fig#1{{Fig.~\ref{#1}}}
\newcommand{\ben}{\begin{eqnarray*}}
\newcommand{\een}{\end{eqnarray*}}

\begin{document}

\title{Long-Range Rapidity Correlations in Heavy-Light Ion Collisions}

\author{Yuri~V.~Kovchegov,\footnote{kovchegov.1@asc.ohio-state.edu}
  Douglas~E.~Wertepny\footnote{wertepny.1@osu.edu}}

\affiliation{Department of Physics, The Ohio State University,
  Columbus, OH 43210, USA}

\begin{abstract}
  We study two-particle long-range rapidity correlations arising in
  the early stages of heavy ion collisions in the saturation/Color
  Glass Condensate framework, assuming for simplicity that one
  colliding nucleus is much larger than the other. We calculate the
  two-gluon production cross section while including all-order
  saturation effects in the heavy nucleus with the lowest-order
  rescattering in the lighter nucleus.  We find four types of
  correlations in the two-gluon production cross section: (i)
  geometric correlations, (ii) HBT correlations accompanied by a
  back-to-back maximum, (iii) away-side correlations, and (iv)
  near-side azimuthal correlations which are long-range in
  rapidity. The geometric correlations (i) are due to the fact that
  nucleons are correlated by simply being confined within the same
  nucleus and may lead to long-range rapidity correlations for the
  produced particles without strong azimuthal angle dependence.
  Somewhat surprisingly, long-range rapidity correlations (iii) and
  (iv) have exactly the same amplitudes along with azimuthal and
  rapidity shapes: one centered around $\Delta \phi =\pi$ with the
  other one centered around $\Delta \phi =0$ (here $\Delta \phi$ is
  the azimuthal angle between the two produced gluons). We thus
  observe that the early-time CGC dynamics in nucleus-nucleus
  collisions generates azimuthal non-flow correlations which are
  qualitatively different from jet correlations by being long-range in
  rapidity. If strong enough, they have the potential of mimicking the
  elliptic (and higher-order even-harmonic) flow in the di-hadron
  correlators: one may need to take them into account in the
  experimental determination of the flow observables.
\end{abstract}

\pacs{25.75.-q, 25.75.Gz, 12.38.Bx, 12.38.Cy}

\maketitle


\section{Introduction}
\label{sec-Intro} 

Long-range rapidity correlations between pairs of hadrons produced at
small azimuthal angles with respect to each other were discovered
recently in heavy ion ($AA$)
\cite{Adams:2005ph,Adare:2008cqb,Alver:2009id,Abelev:2009af},
proton--proton ($pp$) \cite{Khachatryan:2010gv}, and proton--nucleus
($pA$) collisions \cite{CMS:2012qk}. Due to the particular shape of
the corresponding correlation function, with a narrow correlation in
the azimuthal angle $\Delta \phi$ and a wide correlation in
pseudo-rapidity separation $\Delta \eta$, these correlations are often
referred to as the ``ridge''.

There appears to be a consensus in the community that the origin of
these long-range rapidity correlations is in the very early-time
dynamics immediately following the collision. A simple causality
argument demonstrates that a correlation between two hadrons produced
far apart in rapidity may arise only in their common causal past, that
is, in the early stages of the collision
\cite{Dumitru:2008wn,Gavin:2008ev}. However, the detailed dynamical
origin of these ``ridge'' correlations is not completely clear. 

It has been proposed in the literature
\cite{Armesto:2006bv,Armesto:2007ia,Dumitru:2008wn,Gavin:2008ev,Dusling:2009ni,Dumitru:2010iy,Dumitru:2010mv,Kovner:2010xk,Gelis:2008sz}
that the ``ridge'' correlations may arise in the classical gluon field
dynamics of the parton saturation physics/Color Glass Condensate
(CGC). (For reviews of saturation/CGC physics see
\cite{Jalilian-Marian:2005jf,Weigert:2005us,Iancu:2003xm,Gelis:2010nm,KovchegovLevin}.)
Indeed classical gluon fields, which in the McLerran--Venugopalan (MV)
model \cite{McLerran:1994vd,McLerran:1993ka,McLerran:1993ni} dominate
gluon production in heavy ion collisions, do lead to a
rapidity-independent distribution of the produced gluons
\cite{Kovner:1995ja,Kovchegov:1997ke,Krasnitz:2003jw,Blaizot:2010kh}
over rapidity intervals of up to $\Delta y \lesssim 1/\as$, which is
the upper limit of their validity (with $\as$ the strong coupling
constant). Correlations between such classical fields, introduced in
the process of averaging over their color sources, do have a long
range in rapidity
\cite{Kovchegov:1999ep,Dumitru:2008wn,Gavin:2008ev}. Moreover, it was
observed in
\cite{Dumitru:2008wn,Dusling:2009ni,Dumitru:2010mv,Dumitru:2010iy}
that the diagrams giving rise to such rapidity correlations also lead
to a narrow correlation in the azimuthal direction, in qualitative
agreement with the shape of the ``ridge'' correlation.

One has to keep in mind that in heavy ion collisions the early-stage
azimuthal correlation may be washed out by the final state
interactions leading to thermalization of the produced medium and its
hydrodynamic evolution. (Note though, that the rapidity correlation is
not likely to be strongly affected by such late-time dynamics.) It was
argued, however, that the effect of the radial flow in the
hydrodynamic evolution of the quark-gluon plasma (QGP) would be to
(re-)introduce the azimuthal correlations \cite{Gavin:2008ev}.

Another potential complication with the CGC explanation of the
``ridge'' is the fact that rapidity-dependent corrections to classical
gluon fields do become important at rapidity of the order of $\Delta y
\sim 1/\as$. These corrections come in through the non-linear
Balitsky--Kovchegov (BK)
\cite{Balitsky:1996ub,Balitsky:1998ya,Kovchegov:1999yj,Kovchegov:1999ua}
and Jalilian-Marian--Iancu--McLerran--Weigert--Leonidov--Kovner
(JIMWLK)
\cite{Jalilian-Marian:1997dw,Jalilian-Marian:1997gr,Iancu:2001ad,Iancu:2000hn}
evolution equations. Rapidity-dependent corrections are very important
for describing the hadron multiplicity distribution in rapidity,
$dN/dy$, in the CGC framework \cite{ALbacete:2010ad}. As $dN/dy$ does
depend on rapidity rather strongly in RHIC heavy ion data, and also
strongly (albeit less so) in the LHC data, it is very hard to describe
without the rapidity-dependent nonlinear evolution. It is possible
that similar rapidity-dependent corrections may significantly affect
and potentially destroy the long-range structure of the rapidity
correlations due to classical gluon fields of the MV model. Note that
progress on this issue has been made in \cite{Dusling:2009ni},
indicating that inclusion of small-$x$ evolution still leaves the
``ridge'' reasonably flat in rapidity until $\Delta y \sim 1/\as$ when
the correlation disappears.

To elucidate the above questions and concerns, and to improve the
precision of the CGC predictions for the ``ridge''-like correlations,
it is important to be able to calculate the two-particle correlation
in the CGC framework beyond the lowest order. While some works do
consider the role of small-$x$ evolution and multiple rescatterings in
the correlation function
\cite{Dusling:2009ni,Dumitru:2010mv,Kovner:2011pe,Levin:2011fb,KovnerLublinsky12},
most of the phenomenological approaches
\cite{Dusling:2009ni,Dumitru:2010iy,Dusling:2012iga,Dusling:2012wy,Dusling:2012cg}
simply include the saturation effects into the lowest-order
calculation by evolving the unintegrated gluon distributions with the
running-coupling BK (rcBK) nonlinear evolution
\cite{Gardi:2006rp,Kovchegov:2006vj,Balitsky:2006wa}.

The aim of this paper is to begin to analytically include saturation
effects into the two-gluon correlation function in nucleus--nucleus
collisions. Indeed full inclusion analytic of saturation effects
originating in both nuclei would be a very hard problem: even the
single gluon production in the quasi-classical MV limit of $AA$
collisions can be dealt with only numerically at present
\cite{Krasnitz:1999wc,Krasnitz:2003jw,Lappi:2003bi,Blaizot:2010kh}. To
make the problem more tractable we assume that one of the colliding
nuclei is much larger than the other one, such that saturation effects
are important only in interactions with the larger nucleus. In a more
formal language \cite{Kovchegov:1996ty,Kovchegov:1997pc} for the
quasi-classical MV model, we assume that $1 \ll A_1 \ll A_2$, where
$A_1$ and $A_2$ are the atomic numbers of the smaller and larger
nuclei, while $\as^2 \, A_2^{1/3} \sim 1$ and $\as^2 \, A_1^{1/3} \ll
1$. Equivalently one can say that we are interested in production of
gluons with transverse momentum $k_T \gtrsim Q_{s1}$, where $Q_{s1}$
is the saturation scale of the smaller nucleus. We do not impose any
constraints on $k_T$ compared to the saturation scale $Q_{s2}$ of the
larger nucleus ($Q_{s2} \gg Q_{s1} \gg \Lambda_{QCD}$). We stress here
that this setup is not what is usually referred to as the $pA$
collision in the saturation/CGC terminology: since $A_1 \gg 1$, the
two gluon production cross section is dominated by the gluons produced
in interactions of {\sl different} nucleons in the smaller nucleus,
which may also be viewed as a lowest-order saturation correction.

Below we calculate the two-gluon production cross section and the
corresponding correlation function including the all-order saturation
effects in the larger nucleus, while keeping them to the lowest
non-trivial order in the smaller nucleus. The paper is structured as
follows.  In Sec.~\ref{sec:disc} we set up the problem. In the process
of properly defining the correlation function we unexpectedly find a
correlation originating in a simple fact that the nucleons in both
nuclei are confined to within the nuclear radii. The effect is quite
generic, and is not specific to the saturation/CGC approach used in
the calculations in much of this paper: this correlation should be
present in any model of $AA$ collisions which properly takes into
account the geometry of the collision. We refer to these correlations
as the ``geometric correlations''.  We argue that this effect gives a
non-trivial contribution to the correlator even if the two-gluon
production cross section is given by disconnected Feynman diagrams,
i.e., the correlation is not generated dynamically in the
interactions. This correlation may be long-range in rapidity, though
it does not have any non-trivial azimuthal structure.

The two-gluon production cross section in a heavy-light ion collision
is calculated in Sec.~\ref{sec:longrange} in terms of the Wilson line
correlators. We show that the two-gluon production cross section is
related to various correlators of four adjoint Wilson lines: the
adjoint dipole, quadrupole, and a double-trace operator. These
correlators contain both the multiple rescatterings of the
quasi-classical MV approximation, along with the small-$x$ BK/JIMWLK
evolution.

We study the long-range rapidity correlations in
Sec.~\ref{sec:lrrc}. We evaluate the Wilson-line correlators
describing the interaction with the target nucleus using the
quasi-classical MV approximation and the large-$N_c$ limit. By
analyzing the two-gluon correlation function at the lowest non-trivial
order, and in agreement with the earlier calculations in the
literature
\cite{Dusling:2009ni,Dumitru:2010mv,Kovner:2011pe,KovnerLublinsky12},
we find both the near-side ($\Delta \phi =0$) and away-side ($\Delta
\phi =\pi$) long-range rapidity correlations. (Here $\Delta \phi$ is
the azimuthal angle between the momenta of the produced gluons.) The
two correlations are identical in their azimuthal shapes (as functions
of $\Delta \phi$), such that the correlator can be expanded into a
Fourier series in terms of only the even harmonics $\cos 2 \, n \,
\Delta \phi$. We discuss the possibility that such non-flow
correlation may complicate experimental extraction of the contribution
of the true QGP flow to the flow observables $v_{2 n}$. In addition to
the geometric and long-range rapidity correlations, the obtained
two-gluon production cross section contains Hanbury-Brown--Twiss (HBT)
correlations \cite{HanburyBrown:1956pf}. A somewhat peculiar feature
of this HBT correlation is that it is accompanied by an identical
back-to-back peak as well. We conclude in Sec.~\ref{sec:conc} by
stressing the difference between near-side and away-side correlations
calculated here and the mini-jet correlations: the near-side
correlations in this work are long-range in rapidity, while the jet
near-side correlations are local in rapidity.


\section{Correlation function and ``geometric'' correlations}
\label{sec:disc}

\subsection{Definition of the correlator}

Following a standard approach used in experimental analyses of
particle correlations \cite{Eggert:1974ek,Khachatryan:2010gv} the
correlation function can be defined as
\begin{align}\label{corr_def}
  C ( {\bm k}_1, {y}_1, {\bm k}_2, {y}_2 ) = {\cal N} \, \frac{\frac{d
      N_{12}}{d^2 k_1 dy_1 \, d^2 k_2 dy_2}}{\frac{d N}{d^2 k_1 d y_1}
    \, \frac{d N}{d^2 k_2 d y_2}} - 1
\end{align}
where
\begin{align}\label{1part_distr}
  \frac{d N}{d^2 k_1 d y_1} = \frac{1}{\sigma_{inel}} \, \frac{d
    \sigma}{d^2 k_1 d y_1}
\end{align}
and 
\begin{align}\label{2part_distr}
  \frac{d N_{12}}{d^2 k_1 d y_1 \, d^2 k_2 dy_2} =
  \frac{1}{\sigma_{inel}} \, \frac{d \sigma}{d^2 k_1 d y_1 \, d^2 k_2
    dy_2}
\end{align}
are the single- and double-particle multiplicity distributions with
$\sigma_{inel}$ the net inelastic nucleus--nucleus scattering cross
section. Two-dimensional transverse vectors are denoted by ${\bm v} =
(v^x, v^y)$ with their length $v_T \equiv |{\bm v}|$. The
normalization factor $\cal N$ in \eq{corr_def} is fixed by requiring
that the number of particle pairs measured in the same (``real'')
event $N_{12}$ is equal to the number of (``mixed'') pairs with
particles coming from different events $(N)^2$. Here we are interested
in $\Delta \eta$-$\Delta \phi$ correlations with $\Delta \phi = \phi_1
- \phi_2$ the difference between the azimuthal angles $\phi_1, \,
\phi_2$ of the two momenta of the produced particles, and $\delta \eta
= \eta_1 - \eta_2 \approx y_1 - y_2$ the difference between the
pseudo-rapidities $\eta_1, \, \eta_2$ of the two particles. For such a
correlation, with the magnitudes of the transverse momenta $k_{1 \,
  T}$ and $k_{2 \, T}$ constrained to some chosen data bins, the
normalization factor is fixed by
\begin{align}
  \label{eq:norm}
  {\cal N} \int d \phi_1 \, dy_1 \, d \phi_2 \, dy_2 \, \frac{d
    N_{12}}{d^2 k_1 dy_1 \, d^2 k_2 dy_2} = \int d \phi_1 \, d y_1 \,
  \frac{d N}{d^2 k_1 d y_1} \, \int d \phi_2 \, d y_2 \, \frac{d
    N}{d^2 k_2 d y_2}.
\end{align}
(We assume for simplicity that the number of produced particles is
very large, $N \gg 1$, such that $N-1 \approx N$ and $\sigma_{inel}$
is the same in both Eqs. \eqref{1part_distr} and \eqref{2part_distr}
since the production cross section of producing exactly one particle
is negligible.)

Combining Eqs. \eqref{corr_def}, \eqref{1part_distr},
\eqref{2part_distr}, and \eqref{eq:norm} we rewrite the correlation
function in terms of cross sections as
\begin{align}\label{corr_def2}
  C ( {\bm k}_1, {y}_1, {\bm k}_2, {y}_2 ) = \frac{\left[ \int d
      \phi_1 \, d y_1 \, \frac{d \sigma}{d^2 k_1 d y_1} \, \int d
      \phi_2 \, d y_2 \, \frac{d \sigma}{d^2 k_2 d y_2} \right] }{
    \left[ \int d \phi_1 \, dy_1 \, d \phi_2 \, dy_2 \, \frac{d
        \sigma}{d^2 k_1 dy_1 \, d^2 k_2 dy_2} \right]} \ \frac{\frac{d
      \sigma}{d^2 k_1 dy_1 \, d^2 k_2 dy_2} }{\frac{d \sigma}{d^2 k_1
      d y_1} \, \frac{d \sigma}{d^2 k_2 d y_2}} - 1.
\end{align}

As outlined in the Introduction, we take one of the nuclei (the
projectile) to be much smaller than the other, $A_1 \ll A_2$, such
that saturation effects in it are minimal and nucleons in this
(``first'') nucleus interact with the other (``second'') nucleus
independent of each other. At the same time, the projectile nucleus is
still large enough, $A_1 \gg 1$, such that the two-gluon production is
dominated by the gluons generated by the interactions of two different
nucleons from the first (projectile) nucleus with the second
nucleus. (This interaction with the two nucleons in the first nucleus
can be considered a lowest-order saturation effect.) Denote the 1- and
2-nucleon wave functions of the projectile nucleus $A_1$ by $ \Psi_I (
\bm b )$ and $\Psi_{II} ( {\bm b}_1 , {\bm b}_2 )$: they are
normalized such that
\begin{align}\label{wf_norm}
  \int d^2 b \, |\Psi_I ( \bm b )|^2 = A_1, \ \ \ \int d^2 b_1 \, d^2
  b_2 \, |\Psi_{II} ( {\bm b}_1 , {\bm b}_2 )|^2 = A_1 \, (A_1 -1)
  \approx A_1^2.
\end{align}

\begin{figure}[h]
  \includegraphics[width=5cm]{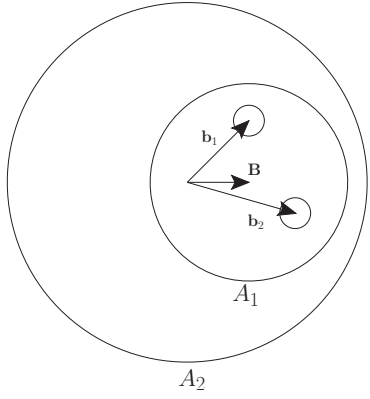}
  \caption{Transverse plane geometry of the two-particle production in
    the collision of a smaller projectile nucleus ($A_1$) with a
    larger target nucleus ($A_2$). The two smaller circles represent
    two nucleons in the nucleus $A_1$ (see text for details).}
\label{nucl_geom} 
\end{figure}

With the help of these wave functions the single- and double-gluon
production cross sections can be written as
\begin{subequations}\label{NNA}
\begin{align}
  \frac{d \sigma}{d^2 k \, d y} & = \int d^2 B \, d^2 b \, | \Psi_I (
  \bm B - \bm b) |^2 \,
  \left\langle \frac{d \sigma^{p A_2}}{d^2 k \, d y \, d^2 b} \right\rangle 
  \label{pA1} \\
  \frac{d \sigma}{d^2 k_1 dy_1 \, d^2 k_2 dy_2} & = \int d^2 B \, d^2
  b_1 \, d^2 b_2 \, |\Psi_{II} ( \bm B - \bm b_1 , \bm B - \bm b_2 )
  |^2 \, \left\langle \frac{d \sigma^{p A_2}}{d^2 k_1 dy_1 d^2 b_1} \,
    \frac{d \sigma^{p A_2}}{d^2 k_2 dy_2 d^2 b_2}
  \right\rangle \label{pA2}
\end{align}
\end{subequations}
where $\bm B$ is the impact parameter between the two nuclei and $\bm
b$, $\bm b_1$, $\bm b_2$ the transverse positions of the nucleons in
the projectile nucleus, all measured with respect to the center of the
second (target) nucleus, as shown in \fig{nucl_geom} for the two-gluon
production process. (Transverse vector $\bm b$ labels the position of
the incoming nucleon in the single gluon production case, while $\bm
b_1$ and $\bm b_2$ label positions of projectile nucleons for
two-gluon production.)  Here
\begin{align}
 \nonumber
 \frac{d \sigma^{pA_2}}{d^2 k \, d y \, d^2 b}
\end{align}
is the cross section for the gluon production (with fixed transverse
momentum $\bm k$, rapidity $y$, and transverse position $\bm b$) in
the collision of a nucleon ($p$) with the target nucleus. The angle
brackets $\langle \ldots \rangle$ in \eqref{NNA} denote averaging in
the target nucleus wave function along with summation over all the
nucleons in the target nucleus
\cite{McLerran:1994vd,McLerran:1993ka,McLerran:1993ni,Kovchegov:1996ty,Jalilian-Marian:1997dw,Jalilian-Marian:1997gr,Iancu:2001ad,Iancu:2000hn}.
Eqs.~\eqref{NNA} can be used in \eq{corr_def2} to give us the
two-gluon correlation function in the heavy-light ion collision
\begin{align}
  \label{eq:Cmain}
  C ( {\bm k}_1, {y}_1, {\bm k}_2, {y}_2) & = \frac{\left[\int d^2 B
      \, d^2 b_1 \, d \phi_1 \, dy_1 \, |\Psi_I ({\bm B} - {\bm
        b}_1)|^2 \, \langle \frac{d \sigma^{p A_2}}{d^2 k_1 \, dy_1 \,
        d^2 b_1} \rangle \right] \, \left[ \int d^2 B \, d^2 b_2 \, d
      \phi_2 \, dy_2 \, |\Psi_I ({\bm B} - {\bm b}_2)|^2 \, \langle
      \frac{d \sigma^{p A_2}}{d^2 k_2 \, dy_2 \, d^2 b_2} \rangle
    \right]}{\left[ \int d^2 B \, d^2 b_1 \, d^2 b_2 \, d \phi_1 \,
      dy_1 \, d \phi_2 \, dy_2 \, |\Psi_{II} ( \bm B - \bm b_1, \bm B
      - \bm b_2 )|^2 \, \left\langle \frac{d \sigma^{p A_2}}{d^2 k_1
          \, dy_1 \, d^2 b_1} \, \frac{d \sigma^{p A_2}}{d^2 k_2 \,
          dy_2 \, d^2 b_2}
      \right\rangle \right]} \notag \\
  & \times \, \frac{\int d^2 B \, d^2 b_1 \, d^2 b_2 \, |\Psi_{II} (
    \bm B - \bm b_1, \bm B - \bm b_2 )|^2 \, \left\langle \frac{d
        \sigma^{p A_2}}{d^2 k_1 dy_1 d^2 b_1} \, \frac{d \sigma^{p
          A_2}}{d^2 k_2 dy_2 d^2 b_2} \right\rangle}{ \left[ \int d^2
      B \, d^2 b_1 \, |\Psi_I ({\bm B} - {\bm b}_1)|^2 \, \left\langle
        \frac{d \sigma^{p A_2}}{d^2 k_1 \, d y_1 \, d^2 b_1}
      \right\rangle \right] \, \left[ \int d^2 B \, d^2 b_2 \, |\Psi_I
      ({\bm B} - {\bm b}_2)|^2 \, \left\langle \frac{d \sigma^{p
            A_2}}{d^2 k_2 \, d y_2 \, d^2 b_2} \right\rangle \right]}
  -1.
\end{align}

Here we model the nucleus as a bag of independent nucleons, which is a
correct description at the leading order in the atomic number $A$
\cite{Mueller:1989st,McLerran:1994vd,McLerran:1993ka,McLerran:1993ni,Kovchegov:1996ty,Kovchegov:1997pc}. In
such case the single-nucleon light-cone wave function squared is
simply equal to the nuclear profile function in the projectile
nucleus,\footnote{We do not show the spin and isospin indices
  explicitly in the wave functions: in our notation the wave function
  squared is implicitly averaged over all nucleon polarizations, since
  both colliding nuclei are unpolarized.}
\begin{align}\label{nmf}
| \Psi_I (\bm b) |^2 = T_1 ({\bm b}).
\end{align}
(The nuclear profile function for a nucleus with density $\rho ({\bm
  b}, z)$ is defined by the integral over the longitudinal coordinate
$z$,
\begin{align}\label{nmf_def}
T ({\bm b}) = \int\limits_{-\infty}^\infty dz \, \rho ({\bm b}, z),
\end{align}
such that for a spherical nucleus of radius $R$ and constant density
$\rho$ it is $T({\bm b} ) = 2 \, \rho \, \sqrt{R^2 - b^2}$.)

Without any loss of generality we can perform the integral over $\bm
B$ in \eq{pA1} obtaining
\begin{align}\label{pA}
  \frac{d \sigma}{d^2 k \, d y} = A_1 \, \int d^2 b \, \left\langle
    \frac{d \sigma^{p A_2}}{d^2 k \, d y \, d^2 b} \right\rangle.
\end{align}

For a sufficiently large projectile nucleus, $A_1 \gg 1$, one can
assume that the two-nucleon wave function can be
factorized,
\begin{align}
  \Psi_{II} ( {\bm b}_1 , {\bm b}_2 ) = \Psi_I ({\bm b}_1) \, \Psi_I
  ({\bm b}_2),
\end{align}
such that, with the help of \eq{nmf} we can write
\begin{align}\label{factorization2}
  |\Psi_{II} ( {\bm b}_1 , {\bm b}_2 )|^2 = T_1 ({\bm b}_1) \, T_1
  ({\bm b}_2).
\end{align}
Using this in \eq{pA2} we get
\begin{align}\label{p2A}
  \frac{d \sigma}{d^2 k_1 dy_1 \, d^2 k_2 dy_2} = \int d^2 B \, d^2
  b_1 \, d^2 b_2 \, T_1 ( \bm B - \bm b_1) \, T_1 (\bm B - \bm b_2 )
  \, \left\langle \frac{d \sigma^{p A_2}}{d^2 k_1 dy_1 d^2 b_1} \,
    \frac{d \sigma^{p A_2}}{d^2 k_2 dy_2 d^2 b_2} \right\rangle
\end{align}

Substituting Eqs.~\eqref{pA} and \eqref{p2A} into \eq{eq:Cmain} yields
\begin{align}\label{corr_main}
  C ( {\bm k}_1, {y}_1, {\bm k}_2, {y}_2 ) = \frac{\left[\int d^2 b_1
      \, d \phi_1 \, d y_1 \, \langle \frac{d \sigma^{p A_2}}{d^2 k_1
        \, dy_1 \, d^2 b_1} \rangle \right] \, \left[\int d^2 b_2 \, d
      \phi_2 \, d y_2 \, \langle \frac{d \sigma^{p A_2}}{d^2 k_2 \,
        dy_2 \, d^2 b_2} \rangle \right]}{\left[ \int d^2 B \, d^2 b_1
      \, d^2 b_2 \, d \phi_1 \, d y_1 \, d \phi_2 \, d y_2 \, T_1 (
      \bm B - \bm b_1) \, T_1 (\bm B - \bm b_2 ) \, \left\langle
        \frac{d \sigma^{p A_2}}{d^2 k_1 \, dy_1 \, d^2 b_1} \, \frac{d
          \sigma^{p A_2}}{d^2 k_2 \,
          dy_2 \, d^2 b_2} \right\rangle \right]} \notag \\
  \times \, \frac{\int d^2 B \, d^2 b_1 \, d^2 b_2 \, T_1 ( \bm B -
    \bm b_1) \, T_1 (\bm B - \bm b_2 ) \, \left\langle \frac{d
        \sigma^{p A_2}}{d^2 k_1 dy_1 d^2 b_1} \, \frac{d \sigma^{p
          A_2}}{d^2 k_2 dy_2 d^2 b_2} \right\rangle}{ \int d^2 b_1 \,
    \left\langle \frac{d \sigma^{p A_2}}{d^2 k_1 \, d y_1 \, d^2 b_1}
    \right\rangle \, \int d^2 b_2 \, \left\langle \frac{d \sigma^{p
          A_2}}{d^2 k_2 \, d y_2 \, d^2 b_2} \right\rangle} -1.
\end{align}
To complete the calculation one needs the single- and double-gluon
production cross sections which, when used in \eq{corr_main}, would
give us the correlation function. Before we proceed to construct them,
let us study a simple example elucidating the nature of one of the
correlation types contained in correlator \eqref{corr_main}.


\subsection{Geometric correlations}

Let us consider the simplest possible example of particle (gluon)
production mechanism where the interaction of the two nucleons in the
first nucleus with the second nucleus in \eq{p2A} factorizes,
\begin{align}\label{int_fact}
  \left\langle \frac{d \sigma^{p A_2}}{d^2 k_1 dy_1 d^2 b_1} \,
    \frac{d \sigma^{p A_2}}{d^2 k_2 dy_2 d^2 b_2} \right\rangle
  \approx \left\langle \frac{d \sigma^{p A_2}}{d^2 k_1 dy_1 d^2 b_1}
  \right\rangle \, \left\langle \frac{d \sigma^{p A_2}}{d^2 k_2 dy_2
      d^2 b_2} \right\rangle.
\end{align}
This contribution comes from the disconnected Feynman diagrams and is
usually identified as the uncorrelated part of the two-gluon
production cross section. However, it is clear that substituting
\eq{int_fact} into \eq{corr_main} does not reduce the correlation
function to zero: instead one gets
\begin{align}\label{corr_fact}
  C ( {\bm k}_1, {y}_1, {\bm k}_2, {y}_2 ) = \frac{\left[\int d^2 b_1
      \, d \phi_1 \, d y_1 \, \langle \frac{d \sigma^{p A_2}}{d^2 k_1
        \, dy_1 \, d^2 b_1} \rangle \right] \, \left[\int d^2 b_2 \, d
      \phi_2 \, d y_2 \, \langle \frac{d \sigma^{p A_2}}{d^2 k_2 \,
        dy_2 \, d^2 b_2} \rangle \right]}{\left[ \int d^2 B \, d^2 b_1
      \, d^2 b_2 \, d \phi_1 \, d y_1 \, d \phi_2 \, d y_2 \, T_1 (
      \bm B - \bm b_1) \, T_1 (\bm B - \bm b_2 ) \, \left\langle
        \frac{d \sigma^{p A_2}}{d^2 k_1 dy_1 d^2 b_1} \right\rangle \,
      \left\langle \frac{d \sigma^{p A_2}}{d^2 k_2 dy_2 d^2 b_2}
      \right\rangle \right]} \notag \\ \times \, \frac{\int d^2 B \,
    d^2 b_1 \, d^2 b_2 \, T_1 ( \bm B - \bm b_1) \, T_1 (\bm B - \bm
    b_2 ) \, \left\langle \frac{d \sigma^{p A_2}}{d^2 k_1 dy_1 d^2
        b_1} \right\rangle \, \left\langle \frac{d \sigma^{p A_2}}{d^2
        k_2 dy_2 d^2 b_2} \right\rangle}{ \int d^2 b_1 \, \left\langle
      \frac{d \sigma^{p A_2}}{d^2 k_1 \, d y_1 \, d^2 b_1}
    \right\rangle \, \int d^2 b_2 \, \left\langle \frac{d \sigma^{p
          A_2}}{d^2 k_2 \, d y_2 \, d^2 b_2} \right\rangle} -1,
\end{align}
which, in general, could be non-zero.

Certainly if the $\bm b$-dependence factorizes from the rapidity and
azimuthal dependence in the cross section
\begin{align}\label{gl_prod}
  \left\langle \frac{d \sigma^{pA_2}}{d^2 k \, d y \, d^2 b}
  \right\rangle
\end{align}
then the correlation function \eqref{corr_fact} is zero: however, such
factorization is not always the case. For gluon production in the
saturation framework, the cross section is a complicated function of
$k_T/Q_s ({\bm b}, y)$, which means it is not in a factorized form and
thus the correlator \eqref{corr_fact} is not zero. Note that in the MV
model (which does not contain the small-$x$ evolution), gluon
production is rapidity-independent, and, if one neglects the
dependence of the gluon production cross section on the angle between
$\bm k$ and $\bm b$, the correlator \eqref{corr_fact} becomes
zero. (Dependence of gluon production cross section on the collision
geometry in the MV approximation is not very strong, peaking at
non-perturbatively low momenta \cite{Teaney:2002kn}.)

For the general case in \eq{corr_fact} we observe a possible
non-trivial correlation in the two-gluon production described by {\sl
  disconnected} Feynman diagrams.  If the gluon production cross
section \eqref{gl_prod} is a slowly varying (but not constant)
function of rapidity, as is the case in the saturation/CGC framework
near mid-rapidity, this correlation would be long-range in
rapidity. The origin of this correlation is somewhat peculiar: even
though the two-nucleon wave function in \eq{factorization2} is
factorized and, hence, represents uncorrelated nucleons, these two
nucleons are correlated by the simple fact of being parts of the same
bound state, the projectile nucleus. In other words, the probability
of finding two nucleons at the impact parameters $\bm b_1$ and $\bm
b_2$ is proportional to
\begin{align}
  \label{eq:prob}
  \sim \, \int d^2 B \, T_1 ( \bm B - \bm b_1) \, T_1 (\bm B - \bm b_2
  )
\end{align}
and is not a product of two independent probabilities after all impact
parameters $\bm B$ of the incoming nucleus are integrated over: this
is a correlation. Note also that the presence of real wave-function
correlations, that is, non-factorizable corrections to the
right-hand-side of \eq{factorization2}, would also lead to some
nontrivial two-particle correlations in \eq{corr_fact}.

If we define the correlation function at the fixed nuclear impact
parameter $\bm B$ by not integrating over $\bm B$ in Eqs.~\eqref{NNA}
and using the result in \eq{corr_def2}, we get
\begin{align}
  \label{eq:CfixedB}
  C ( {\bm k}_1, {y}_1, {\bm k}_2, {y}_2 ; {\bm B} ) =
  \frac{\left[\int d^2 b_1 \, d \phi_1 \, dy_1 \, |\Psi_I ({\bm B} -
      {\bm b}_1)|^2 \, \langle \frac{d \sigma^{p A_2}}{d^2 k_1 \, dy_1
        \, d^2 b_1} \rangle \right] \, \left[\int d^2 b_2 \, d \phi_2
      \, dy_2 \, |\Psi_I ({\bm B} - {\bm b}_2)|^2 \, \langle \frac{d
        \sigma^{p A_2}}{d^2 k_2 \, dy_2 \, d^2 b_2} \rangle
    \right]}{\left[ \int d^2 b_1 \, d^2 b_2 \, d \phi_1 \, dy_1 \, d
      \phi_2 \, dy_2 \, |\Psi_{II} ( \bm B - \bm b_1, \bm B - \bm b_2
      )|^2 \, \langle \frac{d \sigma^{p A_2}}{d^2 k_1 \, dy_1 \, d^2
        b_1} \frac{d
        \sigma^{p A_2}}{d^2 k_2 \, dy_2 \, d^2 b_2} \rangle \right]} \notag \\
  \times \, \frac{\int d^2 b_1 \, d^2 b_2 \, |\Psi_{II} ( \bm B - \bm
    b_1, \bm B - \bm b_2 )|^2 \, \left\langle \frac{d \sigma^{p
          A_2}}{d^2 k_1 dy_1 d^2 b_1} \, \frac{d \sigma^{p A_2}}{d^2
        k_2 dy_2 d^2 b_2} \right\rangle}{ \int d^2 b_1 \, |\Psi_I
    ({\bm B} - {\bm b}_1)|^2 \, \left\langle \frac{d \sigma^{p
          A_2}}{d^2 k_1 \, d y_1 \, d^2 b_1} \right\rangle \, \int d^2
    b_2 \, |\Psi_I ({\bm B} - {\bm b}_2)|^2 \, \left\langle \frac{d
        \sigma^{p A_2}}{d^2 k_2 \, d y_2 \, d^2 b_2} \right\rangle}
  -1,
\end{align}
One can see that this fixed-impact parameter correlation function in
\eq{eq:CfixedB} is zero, $C ({\bm B}) =0$, for the factorized wave
function from \eq{factorization2} {\sl and} for disconnected-diagram
interactions from \eq{int_fact}.  Thus di-gluon correlations due to
our ``geometric'' correlation mechanism seem to also disappear when
the impact parameter is fixed exactly. However, such precise
determination of the impact parameter is impossible in an experimental
analysis, where one is able to fix the collision centrality $|{\bm
  B}|$ in a certain interval, but one can not fix the direction of
$\bm B$. The integration over any range of the impact parameter $|{\bm
  B}|$ (or the integration over the angles of $\bm B$ keeping $|{\bm
  B}|$ constant) is likely to introduce these geometric correlations,
as follows from \eq{corr_main}. Note that the presence of non-trivial
wave function correlations, i.e., correlations beyond the
factorization approximation in \eq{factorization2}, may also lead to
correlations which may survive in \eq{eq:CfixedB} even for a fixed
impact parameter $\bm B$ and uncorrelated interactions
\eqref{int_fact}.

Despite its simplicity, the non-vanishing correlation in
\eq{corr_fact} is one of the main results of this work. In the
saturation/CGC framework it may lead to long-range rapidity
correlations similar to the observed ``ridge'' correlation. Indeed
azimuthal correlations are missing in \eq{corr_fact}: such
correlations may be formed in heavy ion collisions due to radial flow,
as was argued in \cite{Gavin:2008ev}. Indeed a lot more work is needed
to compare this result with experiment.


\section{Two-gluon production and correlations}
\label{sec:longrange}

In this Section we are going to calculate the two-gluon production
cross section for the heavy--light ion collisions working in the
saturation/CGC framework. As described above we assume that $A_2 \gg
A_1 \gg 1$, such that the saturation effects are resummed to all
orders only in the target nucleus with atomic number $A_2$. While the
saturation effects are not very important in the projectile nucleus
with the atomic number $A_1$, the fact that $A_1 \gg 1$ implies that
the two gluons are predominantly produced in collisions of different
nucleons in the projectile nucleus with the target nucleus.

According to the standard technique
\cite{Jalilian-Marian:2005jf,KovchegovLevin}, the calculation will
proceed using the light-cone perturbation theory (LCPT)
\cite{Lepage:1980fj,Brodsky:1983gc} to construct the gluon wave
functions in the incoming projectile nucleus. The production cross
section will be obtained by convoluting the square of this light-cone
wave function with the interaction with the target nucleus, which will
be described by Wilson lines along the light cone
\cite{Balitsky:1996ub,Weigert:2005us}, which include the saturation
effects in the form of both the Glauber--Mueller multiple
rescatterings \cite{Mueller:1989st} and the BK/JIMWLK evolution
equations.

\begin{figure}[htb]
  \includegraphics[width=0.8 \textwidth]{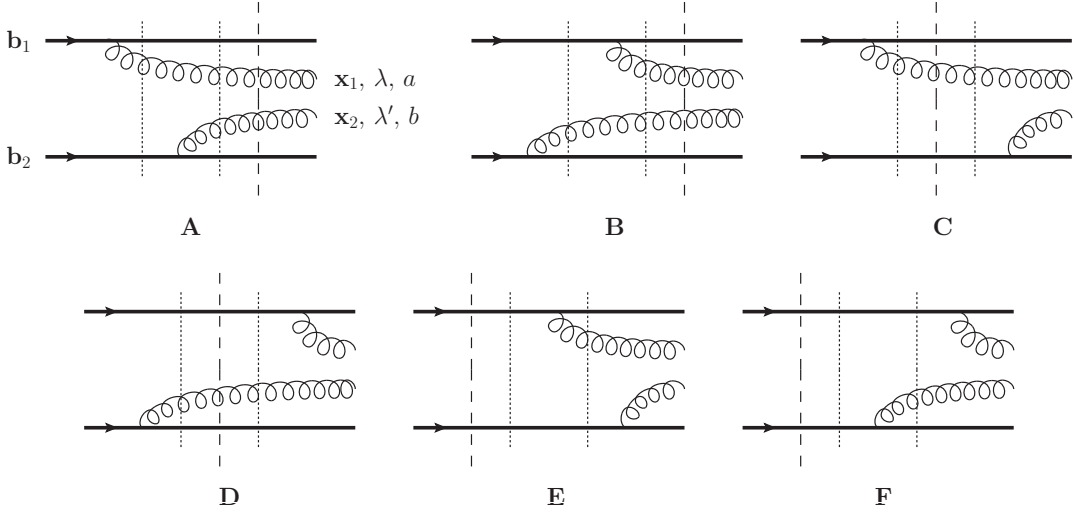}
  \caption{Diagrams describing the two-gluon production in the
    heavy-light ion collision. Two horizontal solid lines denote
    valence quarks inside the two nucleons in the projectile
    nucleus. Vertical dashed line denotes the interaction with the
    target nucleus, while the vertical dotted lines denote
    intermediate states.}
\label{processes} 
\end{figure}

The diagrams contributing to the scattering amplitude of the two-gluon
production in the collision of two nucleons from the projectile
nucleus with the target nucleus are shown in \fig{processes}. There
the two nucleons in the incoming nucleus are denoted by horizontal
solid lines representing two quarks in the wave functions of those
nucleons: throughout this work we will model the nucleons by a single
valence quark each. (Generalization of our results to a more realistic
nucleon wave function is straightforward.) Interaction with the
target, which happens over a time-scale much shorter than the time
scale of preparing gluons in the wave functions, is denoted by a
vertical dashed line.

\subsection{Light cone wave functions of the gluons}

An essential part of the calculation is finding the light cone wave
functions of the two interacting nucleons generating gluons in the
incoming nucleus. The diagrams in \fig{processes} include vertical
dotted lines denoting the intermediate states contributing light-cone
energy denominators to the wave functions. The energy denominators,
and hence the wave functions of the two nucleons, factorize. While
this is obvious in diagrams C and D in \fig{processes}, we will
explain this factorization for other diagrams in a little more detail
below.

We assume that the projectile nucleus is moving along the light cone
``+'' direction and the nucleons carry very large $p^+ = (p^0 +
p^3)/\sqrt{2}$ momenta. At the same time, the gluons carry light cone
momenta $k_1^+$ and $k_2^+$ such that $p^+ \gg k_1^+, \, k_2^+$. The
gluons will be produced with rapidities far away from the
fragmentation regions of the colliding nuclei. The small light cone
momenta of the gluons make them dominate the energy denominators,
generating a much larger contribution than that of the (valence)
quarks \cite{Mueller:1994rr}.

The only difference between the diagrams A and B in \fig{processes} is
the ordering of the emitted gluons. This will only affect the energy
denominators that result from the light-cone perturbation theory rules
\cite{Lepage:1980fj,Brodsky:1983gc}. When these two diagrams are added
together these energy denominators factorize. Denoting the light cone
energies of the two gluons in diagrams A and B by $E_1$ and $E_2$ such
that $E_1 = k_{1 \, T}^2 / (2 \, k_1^+)$ and $E_2 = k_{2 \, T}^2 / (2
\, k_2^+)$ we see that adding the contributions of the energy
denominators in those graphs gives
\begin{align}
  \label{eq:denominators}
  \frac{1}{E_1} \, \frac{1}{E_1 + E_2} + \frac{1}{E_2} \, \frac{1}{E_1
    + E_2} = \frac{1}{E_1} \, \frac{1}{E_2}.
\end{align}
The result factorizes into the terms describing one or another
gluon. Analysis of the diagrams E and F leads to the same
factorization of energy denominators \cite{Kovchegov:2001sc}.

The two-gluon light cone wave function for diagrams A+B and E+F is
\begin{align}
   \label{wavefunction}
   \psi^{ab} \left( {\bm x}_1 , {\bm b}_1 , \lambda ; {\bm x}_2 , {\bm
       b}_2 , \lambda' \right) = \psi^a ({\bm x}_1 , {\bm b}_1 ,
   \lambda) \, \psi^b ({\bm x}_2 , {\bm b}_2 , \lambda') = -
   \frac{g^2}{\pi^2} \; (t^a)_1 \, (t^b)_2 \, \frac{ ( {\bm x}_1 -
     {\bm b}_1) \cdot {\bm \epsilon}_\lambda^*}{ |{\bm x}_1 - {\bm
       b}_1 |^2 } \; \frac{ ( {\bm x}_2 - {\bm b}_2 ) \cdot {\bm
       \epsilon}_{\lambda'}^*} { | {\bm x}_2 - {\bm b}_2 |^2 }
\end{align}
where $g$ is the QCD coupling constant and the polarization vector is
${\bm \epsilon}_\lambda = -(1/\sqrt{2}) (\lambda, i)$
\cite{Lepage:1980fj,Brodsky:1983gc}. The factors of $(t^a)_i$ are the
SU$(N_c)$ generators in the fundamental representation with $N_c$ the
number of quark colors, while the subscript $i=1,2$ denotes the
associated nucleons. The single nucleon's soft gluon wave function is
\cite{Mueller:1994rr}
\begin{align}
  \label{eq:1nucwf}
  \psi^a ({\bm x} , {\bm b} , \lambda) = \frac{i \, g}{\pi} \, t^a \,
  \frac{ ( {\bm x} - {\bm b}) \cdot {\bm \epsilon}_\lambda^*}{ |{\bm
      x} - {\bm b} |^2 }.
\end{align}
The diagrams C+D have a wave function different from \eq{wavefunction}
by a minus sign. The minus sign difference will be absorbed in the
interaction term.


\subsection{Single gluon production}
\label{sec:single}

Our calculation for the two-gluon production cross section will be
closely following that for the single gluon production in a collision
of a single nucleon with a nucleus (a $pA$ collision). For this
reason, and also because we need the single inclusive gluon production
cross section to construct the correlator \eqref{corr_main}, let us
briefly review the results.

\begin{figure}[htb]
  \includegraphics[width=0.5 \textwidth]{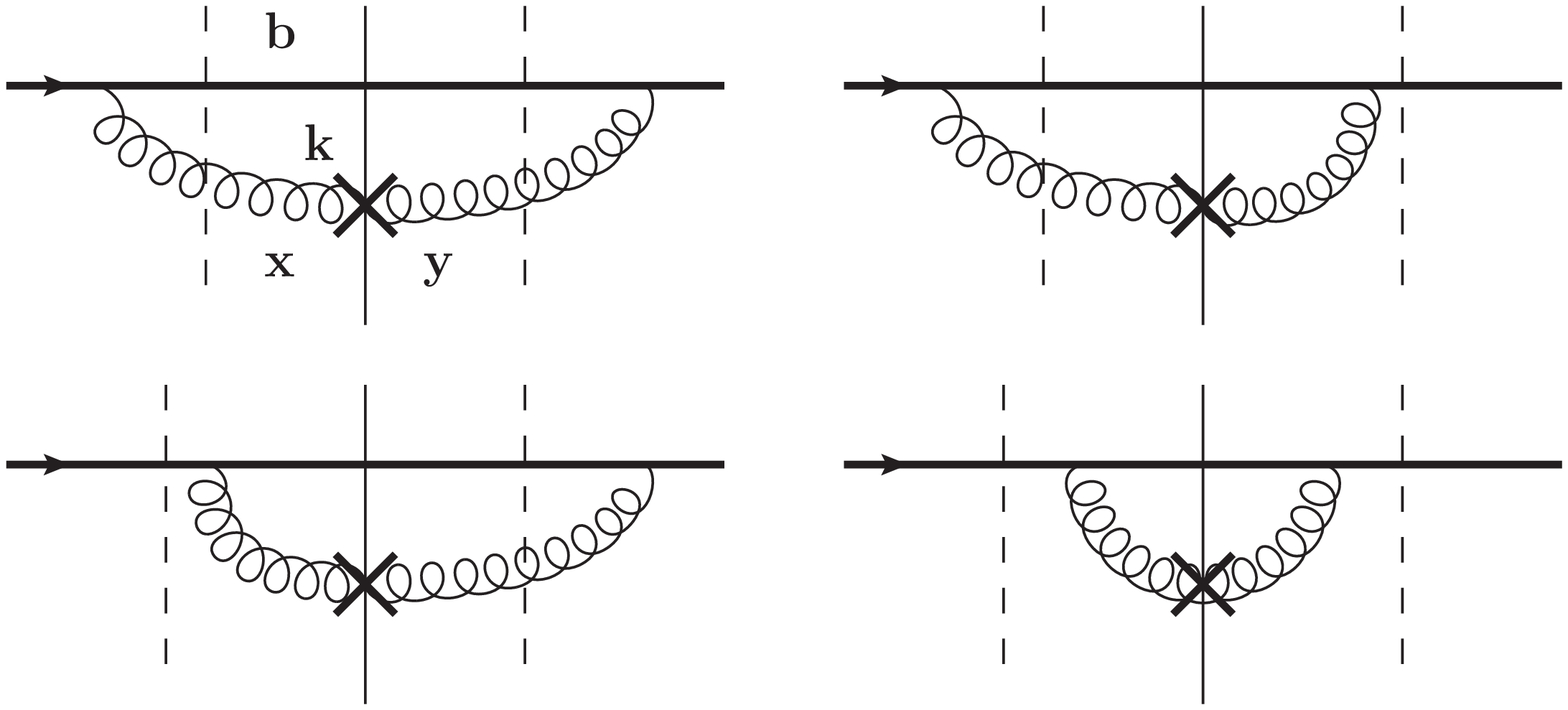}
  \caption{Diagrams contributing to the square of the scattering
    amplitude for the single gluon production in $pA$ collisions. The
    cross denotes the measured produced gluon. }
\label{1Gincl} 
\end{figure}

The single-gluon inclusive production cross section in $pA$ collisions
is illustrated in \fig{1Gincl}, which shows the square of the
scattering amplitude. The cross denotes the measured gluon. The
vertical solid straight line in \fig{1Gincl} is the final-state cut,
while the vertical dashed lines denote interactions with the target
just like in \fig{processes}.

The gluon production cross section can be written as
\cite{Kovchegov:1998bi,Kopeliovich:1998nw,Dumitru:2001ux,Blaizot:2004wu}
(with the projectile proton represented by a single valence quark)
\begin{align}
\label{crosssection1}
\left\langle \frac{d \sigma^{pA_2}}{d^2 k \, dy \, d^2 b}
\right\rangle = \frac{1}{2 \, (2 \pi)^3} \int d^2 x \, d^2 y \; e^{- i
  \, {\bm k} \cdot ({\bm x}-{\bm y})} \, \sum_{\lambda, \, a} \psi^a
({\bm x} , {\bm b} , \lambda) \, \psi^{a \, *} ({\bm y} , {\bm b} ,
\lambda) \, Int_{1} ({\bm x} , {\bm y} , {\bm b})
\end{align}
with the coordinate notation defined in \fig{1Gincl}. The wave
function squared summed over polarizations and colors is
\begin{align}
 \label{wavefunction1}
 \sum_{\lambda, a} \psi^a ({\bm x} , {\bm b} , \lambda) \, \psi^{a \,
   *} ({\bm y} , {\bm b} , \lambda) = \frac{4 \; {\alpha}_s \,
   C_F}{\pi} \, \frac{ {\bm x} - {\bm b}}{ |{\bm x} - {\bm b} |^2 }
 \cdot \frac{ {\bm y} - {\bm b}}{ |{\bm y} - {\bm b} |^2 }
\end{align}
with $C_F = (N_c^2 -1) / (2 N_c)$ the fundamental-representation
Casimir operator, while the interaction with the target is
\begin{align}
 \label{interaction1}
 Int_{1} ({\bm x} , {\bm y} , {\bm b}) = \left\langle
   \frac{1}{N_c^2-1} \; Tr[ U_{{\bm x}} U_{{\bm y}}^\dagger ] \; - \;
   \frac{1}{N_c^2-1} \; Tr[ U_{{\bm x}} U_{{\bm b}}^\dagger ] \; - \;
   \frac{1}{N_c^2-1} \; Tr[ U_{{\bm b}} U_{{\bm y}}^\dagger ] \; + \;
   1 \right\rangle.
\end{align}
The latter object is written in terms of expectation values (in the
target nucleus wave function) of the adjoint Wilson lines along the
$x^+$ light cone:
\begin{align}
  \label{eq:Wline}
  U_{\bm x} = \mbox{P} \exp \left\{ i \, g \,
    \int\limits_{-\infty}^\infty \, d x^+ \, {\cal A}^- (x^+, x^-=0,
    {\bm x}) \right\},
\end{align}
where ${\cal A}^\mu$ is the gluon field of the target in the adjoint
representation.

Defining the adjoint (gluon) dipole $S$-matrix by
\begin{align}
  S_G ( {\bm x}_1 , {\bm x}_2 , y) \equiv \frac{1}{N_c^2-1} \;
  \left\langle Tr[ U_{{\bm x}_1} U_{{\bm x}_2}^\dagger ] \right\rangle
\end{align}
we rewrite \eq{crosssection1} as
\begin{align}
\label{crosssection12}
\left\langle \frac{d \sigma^{pA_2}}{d^2 k \, dy \, d^2 b}
\right\rangle = \frac{\as \, C_F}{4 \, \pi^4} \int d^2 x \, d^2 y \;
e^{- i \, {\bm k} \cdot ({\bm x}-{\bm y})} \, \frac{ {\bm x} - {\bm
    b}}{ |{\bm x} - {\bm b} |^2 } \cdot \frac{ {\bm y} - {\bm b}}{
  |{\bm y} - {\bm b} |^2 } \, \left[ S_G ( {\bm x} , {\bm y}, y ) -
  S_G ( {\bm x} , {\bm b}, y) - S_G ( {\bm b} , {\bm y}, y) + 1
\right].
\end{align}
Using \eq{crosssection12} in \eq{pA} we obtain the gluon production
cross section in the heavy--light ion collision.

The dipole amplitude $S_G$ in \eq{crosssection12} contains the
Glauber-Mueller multiple rescatterings \cite{Mueller:1989st}
(recovered by putting $y=0$ in the argument of $S_G$), along with the
energy dependence included though the BK/JIMWLK evolution equations
\cite{Kovchegov:2001sc,Blaizot:2004wu}: the cross section formula
remains the same in both cases. The former case would correspond to
consistently treating the problem in the MV model
\cite{Kovchegov:1998bi,Kopeliovich:1998nw,Dumitru:2001ux}. The latter
case would include small-$x$ evolution corrections in the rapidity
interval between the produced gluon and the target nucleus: one may
worry that such inclusion would be asymmetric, since the evolution in
the rapidity range between the projectile and the produced target is
not included. A more symmetric treatment of the gluon production
problem in $pA$ collisions with evolution corrections included in all
rapidity intervals is presented in \cite{Kovchegov:2001sc} (see also
\cite{Jalilian-Marian:2005jf,KovchegovLevin} for a more pedagogical
presentation). Inclusion of evolution in the rapidity interval between
the projectile and the produced gluon(s) is beyond the scope of the
present work: it should be possible along the lines of
\cite{Kovchegov:2001sc} though.


\subsection{Two-gluon production with long-range rapidity
  correlations: ``square'' of the single gluon production}

\begin{figure}[h]
  \includegraphics[width= 0.6 \textwidth]{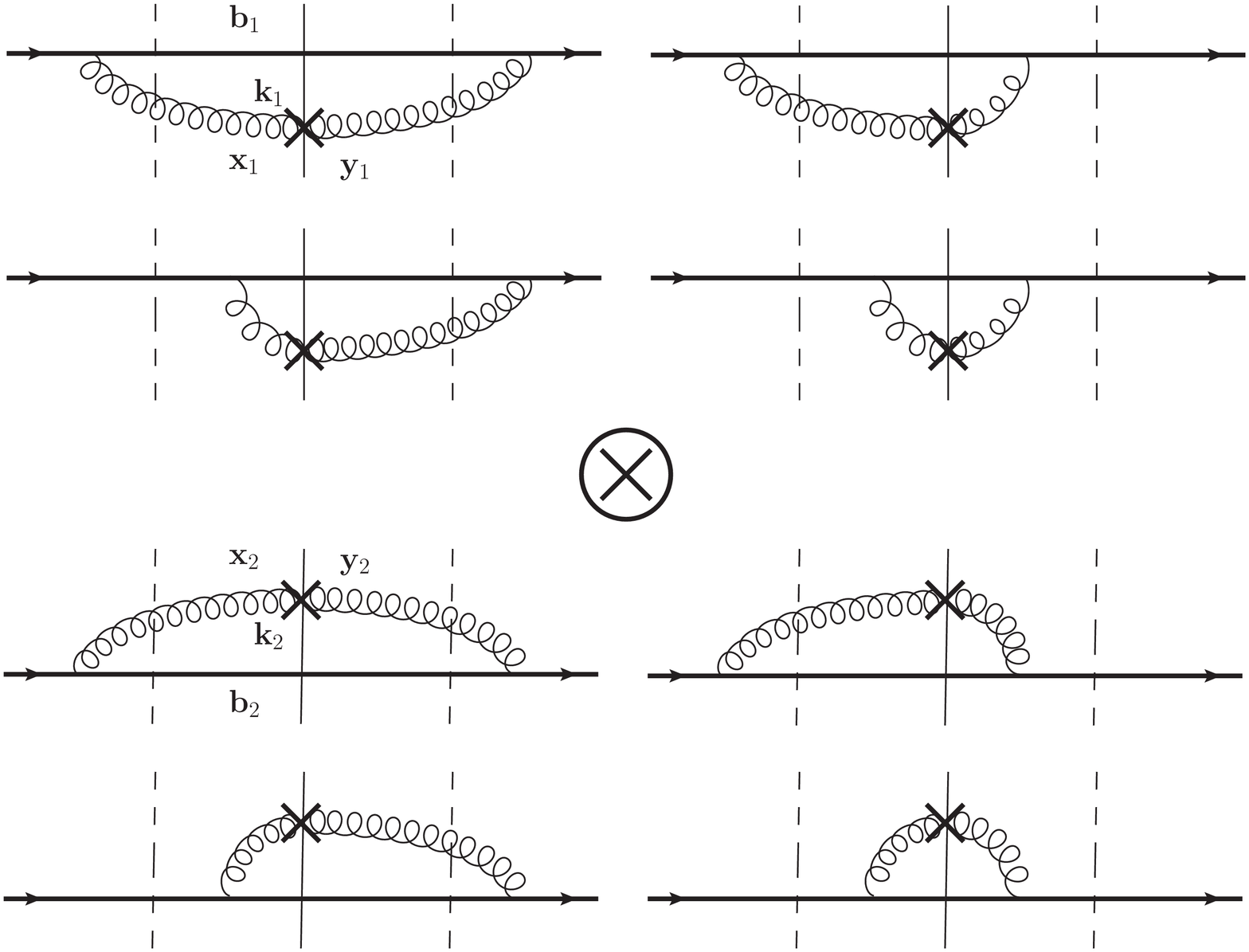}
  \caption{Diagrams contributing to the two-gluon production cross
    section in the heavy--light ion collision. For clarity the
    diagrams are shown as a direct product of gluon production
    processes in collisions of the two interacting nucleons from the
    projectile nucleus with the target nucleus.}
\label{uncrossed} 
\end{figure}

The two-gluon production in heavy-light ion collisions is easily
constructed by analogy to the single-gluon production calculation of
Sec.~\ref{sec:single}. The diagrams contributing to the square of the
scattering amplitude for the double gluon production in heavy--light
ion collisions are shown in \fig{uncrossed}, written as a direct
product of the gluon production processes in the interactions of each
of the nucleons from the projectile nucleus with the target. Just as
in \fig{1Gincl} the vertical dashed lines in \fig{uncrossed} represent
interactions with the target, while the vertical solid line denotes
the final state cut.

The evaluation of the diagrams in \fig{uncrossed} is
straightforward. The two-gluon wave function squared is obtained by
using \eq{wavefunction1} twice, which yields
\begin{align}
 \label{wavefunctionuncrossed}
 \sum_{\lambda, \lambda', a, b} \psi^{ab} ({\bm x}_1, {\bm b_1},
 \lambda; {\bm x}_2, {\bm b_2}, \lambda' ) \, \psi^{a b \, *} ({\bm
   y}_1, {\bm b_1}, \lambda; {\bm y}_2, {\bm b_2}, \lambda' ) \notag
 \\ = \frac{16 \; {\alpha}_s^2 \; C_F^2}{\pi^2} \; \frac{ {\bm x}_1 -
   {\bm b}_1}{ |{\bm x}_1 - {\bm b}_1 |^2 } \cdot \frac{ {\bm y}_1 -
   {\bm b}_1}{ |{\bm y}_1 - {\bm b}_1 |^2 } \ \frac{ {\bm x}_2 - {\bm
     b}_2}{ |{\bm x}_2 - {\bm b}_2 |^2 } \cdot \frac{ {\bm y}_2 - {\bm
     b}_2}{ |{\bm y}_2 - {\bm b}_2 |^2 }
\end{align}
with all the coordinate labels explained in \fig{uncrossed}. 

Interactions with the target are treated similarly. The traces of
Wilson lines in the top and bottom parts of \fig{uncrossed} are
exactly the same as in \eq{interaction1} and appear to factorize,
suggesting absence of dynamically generated correlations. This is not
so. Note, that the averaging over the target is applied to both parts
of \fig{uncrossed} simultaneously: we thus get an averaged product of
traces of Wilson lines,
\begin{align}
 \label{uncrossedinteraction}
 Int_2 ({\bm x}_1, {\bm y}_1, {\bm b}_1; {\bm x}_2, {\bm y}_2, {\bm
   b}_2) & = \left\langle \left( \frac{1}{N_c^2-1} \; Tr[ U_{{\bm
         x}_1} U_{{\bm y}_1}^\dagger ] \; - \; \frac{1}{N_c^2-1} \;
     Tr[ U_{{\bm x}_1} U_{{\bm b}_1}^\dagger ] \; - \;
     \frac{1}{N_c^2-1} \;
     Tr[ U_{{\bm b}_1} U_{{\bm y}_1}^\dagger ] \; + \; 1 \right) \right. \notag \\
 & \times \left. \left( \frac{1}{N_c^2-1} \; Tr[ U_{{\bm x}_2} U_{{\bm
         y}_2}^\dagger ] \; - \; \frac{1}{N_c^2-1} \; Tr[ U_{{\bm
         x}_2} U_{{\bm b}_2}^\dagger ] \; - \; \frac{1}{N_c^2-1} \;
     Tr[ U_{{\bm b}_2} U_{{\bm y}_2}^\dagger ] \; + \; 1 \right)
 \right\rangle,
\end{align}
which does not, in general, factorize into the product of two
separately target-averaged interactions from \eq{interaction1}.

Using \eq{p2A} we write
\begin{align}
  \label{eq:2glue_prod}
  \frac{d \sigma_{square}}{d^2 k_1 dy_1 d^2 k_2 dy_2} & = \frac{1}{[2 (2
    \pi)^3]^2} \int d^2 B \, d^2 b_1 \, d^2 b_2 \, T_1 ({\bm B} - {\bm
    b}_1) \, T_1 ({\bm B} - {\bm b}_2) \, d^2 x_1 \, d^2 y_1 \, d^2
  x_2 \, d^2 y_2 \, e^{- i \; {\bm k}_1 \cdot ({\bm x}_1-{\bm y}_1) -
    i \; {\bm k}_2 \cdot ({\bm x}_2-{\bm y}_2)} \notag \\ & \times \,
  \sum_{\lambda, \lambda', a, b} \psi^{ab} ({\bm x}_1, {\bm b_1},
  \lambda; {\bm x}_2, {\bm b_2}, \lambda' ) \, \psi^{a b \, *} ({\bm
    y}_1, {\bm b_1}, \lambda; {\bm y}_2, {\bm b_2}, \lambda' ) \,
  Int_2 ({\bm x}_1, {\bm y}_1, {\bm b}_1; {\bm x}_2, {\bm y}_2, {\bm
    b}_2),
\end{align}
which, with the help of Eqs.~\eqref{wavefunctionuncrossed} and
\eqref{uncrossedinteraction} becomes (cf. \cite{KovnerLublinsky12})
\begin{align} 
\label{eq:2glue_prod_main} 
\frac{d \sigma_{square}}{d^2 k_1 dy_1 d^2 k_2 dy_2} & = \frac{\as^2 \,
  C_F^2}{16 \, \pi^8} \int d^2 B \, d^2 b_1 \, d^2 b_2 \, T_1 ({\bm B}
- {\bm b}_1) \, T_1 ({\bm B} - {\bm b}_2) \, d^2 x_1 \, d^2 y_1 \, d^2
x_2 \, d^2 y_2 \, e^{- i \; {\bm k}_1 \cdot ({\bm x}_1-{\bm y}_1) - i
  \; {\bm k}_2 \cdot ({\bm x}_2-{\bm y}_2)} \notag \\ & \times \,
\frac{ {\bm x}_1 - {\bm b}_1}{ |{\bm x}_1 - {\bm b}_1 |^2 } \cdot
\frac{ {\bm y}_1 - {\bm b}_1}{ |{\bm y}_1 - {\bm b}_1 |^2 } \ \frac{
  {\bm x}_2 - {\bm b}_2}{ |{\bm x}_2 - {\bm b}_2 |^2 } \cdot \frac{
  {\bm y}_2 - {\bm b}_2}{ |{\bm y}_2 - {\bm b}_2 |^2 } \notag \\
& \times \, \left\langle \left( \frac{1}{N_c^2-1} \; Tr[ U_{{\bm x}_1}
    U_{{\bm y}_1}^\dagger ] \; - \; \frac{1}{N_c^2-1} \; Tr[ U_{{\bm
        x}_1} U_{{\bm b}_1}^\dagger ] \; - \; \frac{1}{N_c^2-1} \;
    Tr[ U_{{\bm b}_1} U_{{\bm y}_1}^\dagger ] \; + \; 1 \right) \right. \notag \\
& \times \left. \left( \frac{1}{N_c^2-1} \; Tr[ U_{{\bm x}_2} U_{{\bm
        y}_2}^\dagger ] \; - \; \frac{1}{N_c^2-1} \; Tr[ U_{{\bm x}_2}
    U_{{\bm b}_2}^\dagger ] \; - \; \frac{1}{N_c^2-1} \; Tr[ U_{{\bm
        b}_2} U_{{\bm y}_2}^\dagger ] \; + \; 1 \right) \right\rangle.
\end{align}
This is the expression for the two-gluon production cross section
coming from the diagrams in \fig{uncrossed}. The interaction with the
target can be evaluated in the quasi-classical multiple rescattering
approximation, as we will show later. The rapidity evolution can be
included using the JIMWLK equation. Note, however, that when the
rapidity difference between the two gluons is sufficiently large,
$|y_1 - y_2| \gtrsim 1/\as$, one has to include the evolution
corrections in the rapidity interval between the produced gluons, such
that the Wilson lines in the two parenthesis in
\eq{eq:2glue_prod_main} should be taken at different rapidities. A
similar effect had to be included in the two-gluon production cross
section in DIS in \cite{JalilianMarian:2004da}. In such regime one
would also need to include evolution corrections in the rapidity
window between the projectile and (at least one of) the produced
gluons. Inclusion of the evolution corrections in terms of the weight
functional $W$ of the JIMWLK evolution equation into the two-gluon
production cross section in nucleus--nucleus collisions was done in
\cite{Dusling:2009ni}. In this work we will limit ourselves to the
quasi-classical regime where no evolution corrections are required and
the cross sections are rapidity-independent.

Note that for scattering on a large nuclear target, if one resums
powers of $\as^2 \, A^{1/3}$, as is the case in the Glauber-Mueller
(GM) rescatterings (and in the BK/JIMWLK evolution for which the GM
rescatterings serve as the initial condition), and if one takes the
large-$N_c$ limit, the expectation values of the traces in
\eq{eq:2glue_prod_main} factorize, such that
\begin{align}
  \label{eq:trace_fact}
  \left\langle Tr[ U_{{\bm x}_1} U_{{\bm y}_1}^\dagger ] \; Tr[
    U_{{\bm x}_2} U_{{\bm y}_2}^\dagger ] \right\rangle
  \bigg|_{\mbox{large}-N_c, \, \mbox{large}-A_2} \approx \left\langle
    Tr[ U_{{\bm x}_1} U_{{\bm y}_1}^\dagger ] \right\rangle \;
  \left\langle Tr[ U_{{\bm x}_2} U_{{\bm y}_2}^\dagger ]
  \right\rangle.
\end{align}
This leads to factorization of \eq{int_fact} being valid in the
large-$N_c$ and large-target-nucleus limit. As shown above, even in
this factorized regime one may obtain a non-trivial correlation
function due to the geometric correlations.


\subsection{Two-gluon production with long-range rapidity
  correlations: ``crossed'' diagrams}

\begin{figure}[h]
  \includegraphics[width= 0.79 \textwidth]{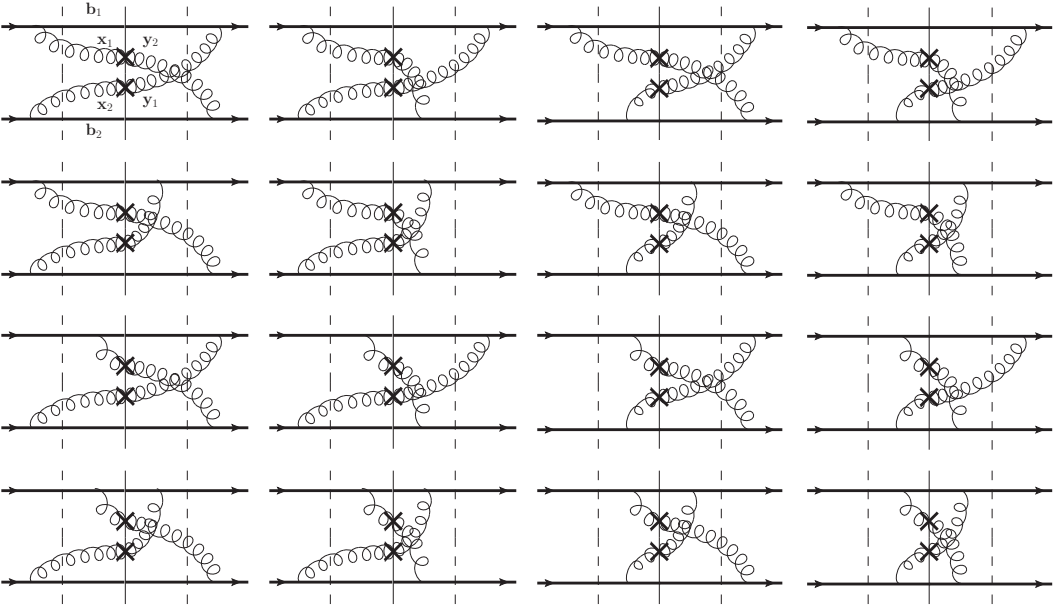}
  \caption{Diagrams contributing to the two-gluon production cross
    section, with the gluon emitted by each nucleon in the amplitude
    absorbed by another nucleon in the complex conjugate
    amplitude. The top cross denotes the gluon with momentum ${\bm
      k}_1$, while the bottom one denotes the gluon with momentum
    ${\bm k}_2$.}
\label{crossed} 
\end{figure}

\begin{figure}[h]
  \includegraphics[width= 0.79 \textwidth]{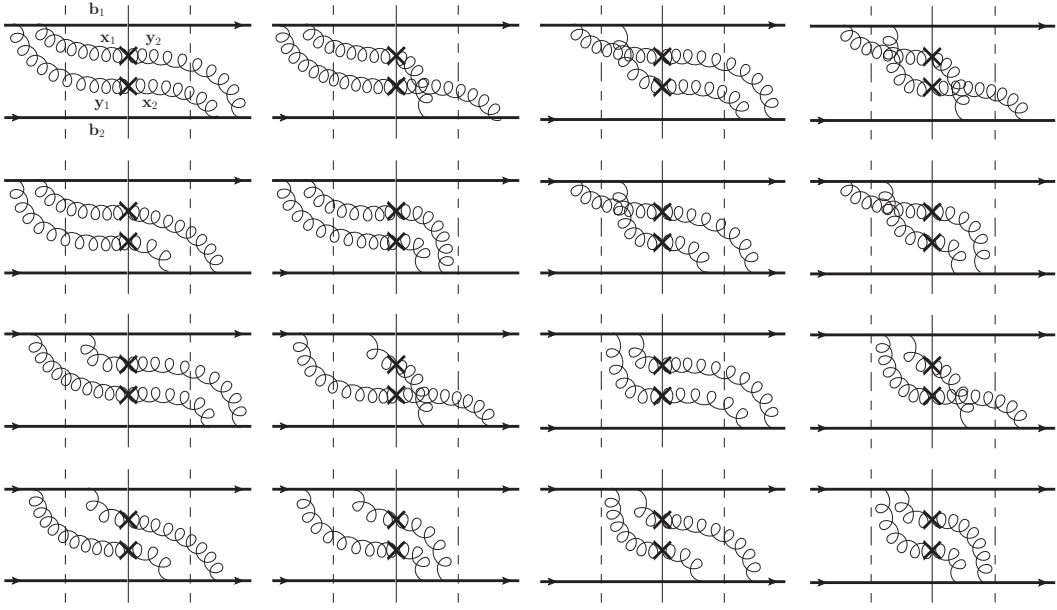}
  \caption{Another set of diagrams contributing to the two-gluon
    production cross section, with the gluon emitted by each nucleon
    in the amplitude absorbed by another nucleon in the complex
    conjugate amplitude. Again the top cross denotes the gluon with
    momentum ${\bm k}_1$, while the bottom one denotes the gluon with
    momentum ${\bm k}_2$. Summation over the different orderings of
    gluon emissions (e.g. which gluon is emitted first or second) is
    implied but is not shown explicitly. }
\label{crossed2} 
\end{figure}

Before we proceed to evaluating the correlations contained in the
cross section \eqref{eq:2glue_prod_main}, let us point out another
contribution to the two-gluon production cross section arising from
squaring the sum of the diagrams in \fig{processes}. When squaring the
diagrams in \fig{processes} it is possible that the gluon emitted by
one nucleon in the amplitude will be absorbed by another nucleon in
the complex conjugate amplitude. The corresponding contributions to
the cross section are shown in Figs.~\ref{crossed} and \ref{crossed2},
where crossing gluon lines do not form a vertex. We will refer to
these diagrams as the ``crossed'' graphs. The diagrams obtained from
those in Figs.~\ref{crossed} and \ref{crossed2} by a mirror reflection
with respect to the cut correspond to the ${\bm b}_1 \leftrightarrow
{\bm b}_2$ interchange, and will be automatically included in the
cross section to be calculated below since it will contain integrals
over all ${\bm b}_1$ and ${\bm b}_2$. The sum over different orderings
of gluon emissions in the (complex conjugate) amplitude of
\fig{crossed2} is implied, but is not shown explicitly.


In evaluating the graphs in Figs.~\ref{crossed} and \ref{crossed2} we
note that there are non-trivial color factors due to the interaction
terms, which we absorb into the wave function. The expression for the
wave function squared is
\begin{align}
 \label{wavefunctioncrossed}
 \frac{16 \; {\alpha}_s^2}{\pi^2} \; \frac{C_F}{2 N_c} \; \frac{ {\bm
     x}_1 - {\bm b}_1}{|{\bm x}_1 - {\bm b}_1|^2} \cdot \frac{{\bm
     y}_2 - {\bm b}_2 }{ |{\bm y}_2 - {\bm b}_2|^2 } \, \frac{{\bm
     x}_2 - {\bm b}_2}{|{\bm x}_2 - {\bm b}_2|^2} \cdot \frac{{\bm
     y}_1 - {\bm b}_1 }{ |{\bm y}_1 - {\bm b}_1|^2}.
\end{align}

Defining the color-quadrupole operator
\cite{Chen:1995pa,JalilianMarian:2004da,Dominguez:2011gc,Dumitru:2011vk,Iancu:2011ns}
\begin{align}\label{quad_def}
  Q ( {\bm x}_1 , {\bm x}_2 , {\bm x}_3 , {\bm x}_4 ) \equiv
  \frac{1}{N_c^2-1} \; \left\langle Tr[ U_{{\bm x}_1} U_{{\bm
        x}_2}^\dagger U_{{\bm x}_3} U_{{\bm x}_4}^\dagger ]
  \right\rangle
\end{align}
after some algebra the interactions with the target in both
\fig{crossed} and \fig{crossed2}
can be shown to be equal to (with the terms ordered in the same way as
the diagrams in Figs.~\ref{crossed}
and \ref{crossed2} and the transverse coordinates defined in the upper
left diagram of each of these two figures)
\begin{align}
 \label{crossedinteraction}
 Int_{crossed} ( {\bm x}_1, {\bm y}_1 , {\bm b}_1 , {\bm x}_2 , {\bm
   y}_2, {\bm b}_2) &= \; Q( {\bm x}_1, {\bm y}_1 , {\bm x}_2 , {\bm
   y}_2 ) - Q( {\bm x}_1, {\bm y}_1 , {\bm x}_2 , {\bm b}_2 ) - Q(
 {\bm x}_1, {\bm y}_1 , {\bm b}_2 , {\bm y}_2 ) + S_G( {\bm x}_1, {\bm
   y}_1)
 \notag \\
 & - Q( {\bm x}_1, {\bm b}_1 , {\bm x}_2 , {\bm y}_2 ) \; + \; Q( {\bm
   x}_1, {\bm b}_1 , {\bm x}_2 , {\bm b}_2 ) + Q( {\bm x}_1, {\bm b}_1
 , {\bm b}_2 , {\bm y}_2 ) - S_G( {\bm x}_1, {\bm b}_1)
 \notag \\
 & - Q( {\bm b}_1, {\bm y}_1 , {\bm x}_2 , {\bm y}_2 ) + Q( {\bm b}_1,
 {\bm y}_1 , {\bm x}_2 , {\bm b}_2 ) + Q( {\bm b}_1, {\bm y}_1 , {\bm
   b}_2 , {\bm y}_2 ) - S_G( {\bm b}_1, {\bm y}_1)
 \notag \\
 & + S_G( {\bm x}_2, {\bm y}_2) - S_G( {\bm x}_2, {\bm b}_2) - S_G(
 {\bm b}_2, {\bm y}_2) + 1.
\end{align}
The only difference between the contributions of the diagrams in
Figs.~\ref{crossed} and \ref{crossed2} is in the exponential factors
for the Fourier transform into the transverse momentum space.

\begin{figure}[hb]
  \includegraphics[width= 0.44 \textwidth]{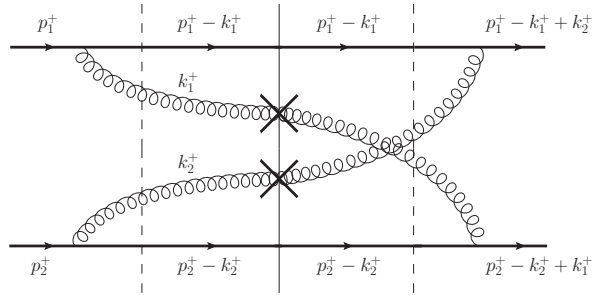}
  \caption{The flow of the ``plus'' momentum component through the
    first diagram in \fig{crossed}.}
\label{hbt_rapidities} 
\end{figure}

The longitudinal momentum flow patterns in Figs.~\ref{crossed} and
\ref{crossed2} are different from that in \fig{uncrossed}.  This is
illustrated in \fig{hbt_rapidities}, which shows the flow of the
``plus'' momentum component through the first diagram in
\fig{crossed}. Note that the change in the ``plus'' momentum component
is negligible in the eikonal interactions with the target considered
here. Requiring that the incoming quark lines carry the same ``plus''
momentum both in the amplitude and in the complex conjugate amplitude,
one would obtain $k_1^+ = k_2^+$ : however such requirement is not
correct. The actual scattering happens between two nuclei, and it is
the momenta of the whole incoming nuclei which have to be equal both
in the amplitude and in the complex conjugate amplitude. Hence for
$k_1^+ \neq k_2^+$ the diagrams like that in \fig{hbt_rapidities}
would only correspond to different redistributions of the projectile
nucleus momentum between the nucleons in it in the amplitude and in
the complex conjugate amplitude without changing the same ``plus''
momentum of the whole nucleus on both sides of the cut. Hence the
$k_1^+ = k_2^+$ condition is not necessary for the ``crossed''
diagrams.

Combining the factors in Eqs.~\eqref{wavefunctioncrossed} and
\eqref{crossedinteraction}, and inserting Fourier transform
exponentials, we arrive at the expression for the two-gluon production
cross section contribution resulting from the ``crossed'' diagrams
from Figs.~\ref{crossed} and \ref{crossed2}
\begin{align}\label{crossed_xsect}
  & \frac{d \sigma_{crossed}}{d^2 k_1 dy_1 d^2 k_2 dy_2} =
  \frac{1}{[2(2 \pi)^3]^2} \, \int d^2 B \, d^2 b_1 \, d^2 b_2 \, T_1
  ({\bm B} - {\bm b}_1) \, T_1 ({\bm B} - {\bm b}_2) \, d^2 x_1 \, d^2
  y_1 \, d^2 x_2 \, d^2 y_2 \notag \\ & \times \, \left[ e^{- i \;
      {\bm k}_1 \cdot ({\bm x}_1-{\bm y}_2) - i \; {\bm k}_2 \cdot
      ({\bm x}_2-{\bm y}_1)} + e^{- i \; {\bm k}_1 \cdot ({\bm
        x}_1-{\bm y}_2) + i \; {\bm k}_2 \cdot ({\bm x}_2-{\bm y}_1)}
  \right] \, \frac{16 \; {\alpha}_s^2}{\pi^2} \, \frac{C_F}{2 N_c} \;
  \frac{ {\bm x}_1 - {\bm b}_1}{|{\bm x}_1 - {\bm b}_1|^2} \cdot
  \frac{{\bm y}_2 - {\bm b}_2 }{ |{\bm y}_2 - {\bm b}_2|^2 } \,
  \frac{{\bm x}_2 - {\bm b}_2}{|{\bm x}_2 - {\bm b}_2|^2} \cdot
  \frac{{\bm y}_1 - {\bm b}_1 }{ |{\bm y}_1 - {\bm b}_1|^2} \notag \\
  & \times \, \bigg[ Q( {\bm x}_1, {\bm y}_1 , {\bm x}_2 , {\bm y}_2 )
  - Q( {\bm x}_1, {\bm y}_1 , {\bm x}_2 , {\bm b}_2 ) - Q( {\bm x}_1,
  {\bm y}_1 , {\bm b}_2 , {\bm y}_2 ) + S_G( {\bm x}_1, {\bm y}_1) -
  Q( {\bm x}_1, {\bm b}_1 , {\bm x}_2 , {\bm y}_2 ) \; + \; Q( {\bm
    x}_1, {\bm b}_1 , {\bm x}_2 , {\bm b}_2 ) \notag \\ & + Q( {\bm
    x}_1, {\bm b}_1 , {\bm b}_2 , {\bm y}_2 ) - S_G( {\bm x}_1, {\bm
    b}_1) - Q( {\bm b}_1, {\bm y}_1 , {\bm x}_2 , {\bm y}_2 ) + Q(
  {\bm b}_1, {\bm y}_1 , {\bm x}_2 , {\bm b}_2 ) + Q( {\bm b}_1, {\bm
    y}_1 , {\bm b}_2 , {\bm y}_2 ) - S_G( {\bm b}_1, {\bm y}_1) + S_G(
  {\bm x}_2, {\bm y}_2)
  \notag \\
  & - S_G( {\bm x}_2, {\bm b}_2) - S_G( {\bm b}_2, {\bm y}_2) +
  1\bigg].
\end{align}
Just like in \eq{eq:2glue_prod_main}, the dipole and quadrupole
scattering amplitudes in \eq{crossed_xsect} can be evaluated either in
the MV model or by using BK and JIMWLK evolution equations. The
quadrupole amplitude evolution equation was derived in the large-$N_c$
limit in \cite{JalilianMarian:2004da}, and beyond the large-$N_c$
limit in \cite{Dominguez:2011gc}. Again, \eq{crossed_xsect} is valid
only as long as the rapidities $y_1$ and $y_2$ of the two produced
gluons are close to each other, $|y_2 - y_1| \lesssim 1/\as$, such
that no small-$x$ evolution corrections need to be included in the
$[y_1, y_2]$ rapidity interval.


\subsection{Two-gluon production with long-range rapidity
  correlations: the net result}

Equations (\ref{eq:2glue_prod_main}) and (\ref{crossed_xsect}), when
combined, give us the two-gluon production cross section in the
heavy-light ion collisions:
\begin{align}\label{eq_all}
  \frac{d \sigma}{d^2 k_1 dy_1 d^2 k_2 dy_2} = \frac{d
    \sigma_{square}}{d^2 k_1 dy_1 d^2 k_2 dy_2} + \frac{d
    \sigma_{crossed}}{d^2 k_1 dy_1 d^2 k_2 dy_2}.
\end{align}
This production cross section is the main formal result of this
work. We analyze the properties of the cross section \eqref{eq_all}
below.


\section{Long-range rapidity correlations: away-side and near-side;
  HBT correlations}
\label{sec:lrrc}

Our goal now is to evaluate the correlations resulting from the
two-gluon production cross section \eqref{eq_all}, that is from the
cross sections in Eqs.~(\ref{eq:2glue_prod_main}) and
(\ref{crossed_xsect}). The first step is to evaluate the interaction
with the target. We will be working in the quasi-classical MV/GM
limit, where the interaction and, hence, the production cross sections
\eqref{eq:2glue_prod_main} and \eqref{crossed_xsect}, are
rapidity-independent.

We begin with the cross section in \eq{eq:2glue_prod_main}. Even in
the quasi-classical limit, evaluation of the interaction in
\eq{uncrossedinteraction} is calculationally intensive (though
conceptually rather straightforward). To simplify the calculation we
will also employ the large-$N_c$ expansion: we assume that the
nucleons in the nuclei are made out of an order-$N_c^2$ valence quarks
(or gluons), such that the saturation scale, which in such case is
proportional to $Q_{s0}^2 \sim \as^2 N_c^2$, is constant in the 't
Hooft's large-$N_c$ limit. In the saturation physics framework such
approximation was used in \cite{JalilianMarian:2004da} for the
quadrupole operator giving a reasonably good approximation to the
exact answer \cite{Dumitru:2011vk}. Similar approximations are
frequently used (albeit, often implicitly) in applications of anti-de
Sitter space/conformal field theory (AdS/CFT) correspondence to
collisions of heavy ions modeled by shock waves
\cite{Grumiller:2008va,Albacete:2008vs,Albacete:2009ji,Chesler:2010bi}.

At the leading order in $1/N_c^2$ expansion of \eq{eq:2glue_prod_main}
the interaction with the target factorizes, as discussed around
\eq{eq:trace_fact}. In addition, the cross section in
\eq{crossed_xsect} is $1/N_c^2$-suppressed (as compared to the leading
term in \eq{eq:2glue_prod_main}) and can be neglected at the leading
order in $N_c$. The correlation function is then given by
\eq{corr_fact} with the single gluon production cross section from
\eq{crosssection12}. In the MV/GM approximation the gluon color dipole
interaction with the target is \cite{Mueller:1989st}
\begin{align}
  \label{eq:SG_GM}
  S_G ({\bm x}_1, {\bm x}_2, y=0) = \exp \left[ -\frac{1}{4} \, |{\bm
      x}_1 - {\bm x}_2|^2 \, Q_{s0}^2 \left( \frac{{\bm x}_1 + {\bm
          x}_2}{2} \right) \, \ln \left(\frac{1}{|{\bm x}_1 - {\bm
          x}_2| \, \Lambda} \right) \right]
\end{align}
with $Q_{s0}$ the rapidity-independent gluon saturation scale in the
quasi-classical limit evaluated at the dipole center-of-mass $({\bm
  x}_1 + {\bm x}_2)/2$ and $\Lambda$ an infrared (IR) cutoff. We see
that the quasi-classical single gluon production cross section is
rapidity-independent and, for unpolarized nuclei and for
perturbatively large $k_T$ \cite{Teaney:2002kn}, is also independent
of the azimuthal angle $\phi$ of the transverse momentum $\bm k$ of
the outgoing gluon. We conclude that at the leading order in $1/N_c^2$
in the quasi-classical approximation the geometric correlations are
almost absent and the correlation function \eqref{corr_fact} is
approximately zero.

Non-trivial correlations can be obtained from \eq{eq:2glue_prod_main}
by expanding the interaction term to the first non-trivial order in
$1/N_c^2$. To this end we write a correlator of two Wilson line traces
as
\begin{align}\label{Ddef}
  \frac{1}{(N_c^2 - 1)^2} \, \langle Tr[ U_{{\bm x}_1} U_{{\bm
      x}_2}^\dagger ] \, Tr[ U_{{\bm x}_3} U_{{\bm x}_4}^\dagger ]
  \rangle = \frac{1}{(N_c^2 - 1)^2} \, \langle Tr[ U_{{\bm x}_1}
  U_{{\bm x}_2}^\dagger ] \rangle \langle Tr[ U_{{\bm x}_3} U_{{\bm
      x}_4}^\dagger ] \rangle + \Delta ( {\bm x}_1 , {\bm x}_2 , {\bm
    x}_3 , {\bm x}_4 ),
\end{align}
where the correction to the factorized expression
\eqref{eq:trace_fact}, denoted by $\Delta$, is order-$1/N_c^2$. (Note
that each adjoint trace in \eq{Ddef} is order-$(N_c^2 -1)$, such that
its left-hand side, as well as the first term on its right-hand side,
are order-one in $N_c$ counting.)

The leading order-$1/N_c^2$ contribution to $\Delta$ is derived in the
Appendix, with the result being
\begin{align}
  \label{Delta_exp}
  \Delta ( {\bm x}_1 , {\bm x}_2 , {\bm x}_3 , {\bm x}_4 ) =
  \frac{(D_3-D_2)^2}{N_c^2} \, & \left[ \frac{e^{D_1}}{D_1-D_2} -
    \frac{2 \, e^{D_1}}{(D_1-D_2)^2} + \frac{e^{D_1}}{D_1-D_3} -
    \frac{2 \,
      e^{D_1}}{(D_1-D_3)^2} \right. \notag \\
  & \left. + \frac{2 \, e^{\frac{1}{2}(D_1+D_2)}}{(D_1-D_2)^2} +
    \frac{2 \, e^{\frac{1}{2}(D_1+D_3)}}{(D_1-D_3)^2} \right] + O
  \left( \frac{1}{N_c^4} \right),
\end{align}
where we have defined
\begin{subequations}\label{Ds}
\begin{align}
  D_1 & = - \frac{Q_{s0}^2}{4} \left[ | \bm x_1 - \bm x_2 |^2 \, \ln \left( \frac{1}{| \bm x_1 - \bm x_2 | \Lambda } \right) + | \bm x_3 - \bm x_4 |^2 \, \ln \left( \frac{1}{| \bm x_3 - \bm x_4 | \Lambda } \right) \right] \\
  D_2 & = - \frac{Q_{s0}^2}{4} \left[ | \bm x_1 - \bm x_3 |^2 \, \ln \left( \frac{1}{| \bm x_1 - \bm x_3 | \Lambda } \right) + | \bm x_2 - \bm x_4 |^2 \, \ln \left( \frac{1}{| \bm x_2 - \bm x_4 | \Lambda } \right) \right] \\
  D_3 & = - \frac{Q_{s0}^2}{4} \left[ | \bm x_1 - \bm x_4 |^2 \, \ln
    \left( \frac{1}{| \bm x_1 - \bm x_4 | \Lambda } \right) + | \bm
    x_2 - \bm x_3 |^2 \, \ln \left( \frac{1}{| \bm x_2 - \bm x_3 |
        \Lambda } \right) \right]
\end{align}
\end{subequations}
assuming, for simplicity, that all the saturation scales are evaluated
at the same impact parameter.

Using \eq{Ddef} in \eq{eq:2glue_prod_main} we see that the correlated
part of the two-gluon production cross section is
\begin{align} 
\label{eq:2glue_prod_delta} 
\frac{d \sigma_{square}^{(corr)}}{d^2 k_1 dy_1 d^2 k_2 dy_2} & = \frac{\as^2 \,
  C_F^2}{16 \, \pi^8} \int d^2 B \, d^2 b_1 \, d^2 b_2 \, T_1 ({\bm B}
- {\bm b}_1) \, T_1 ({\bm B} - {\bm b}_2) \, d^2 x_1 \, d^2 y_1 \, d^2
x_2 \, d^2 y_2 \, e^{- i \; {\bm k}_1 \cdot ({\bm x}_1-{\bm y}_1) - i
  \; {\bm k}_2 \cdot ({\bm x}_2-{\bm y}_2)} \notag \\ & \times \,
\frac{ {\bm x}_1 - {\bm b}_1}{ |{\bm x}_1 - {\bm b}_1 |^2 } \cdot
\frac{ {\bm y}_1 - {\bm b}_1}{ |{\bm y}_1 - {\bm b}_1 |^2 } \ \frac{
  {\bm x}_2 - {\bm b}_2}{ |{\bm x}_2 - {\bm b}_2 |^2 } \cdot \frac{
  {\bm y}_2 - {\bm b}_2}{ |{\bm y}_2 - {\bm b}_2 |^2 } \notag \\
& \times \, \left[ \Delta( {\bm x}_1 , {\bm y}_1 , {\bm x}_2 , {\bm
    y}_2 ) - \Delta ( {\bm x}_1 , {\bm y}_1 , {\bm x}_2 , {\bm b}_2 )
  - \Delta ( {\bm x}_1 , {\bm y}_1 , {\bm b}_2 , {\bm y}_2 ) - \Delta
  ( {\bm x}_1 , {\bm b}_1 , {\bm x}_2 , {\bm y}_2 )
  - \Delta ( {\bm b}_1 , {\bm y}_1 , {\bm x}_2 , {\bm y}_2 ) \right. \notag \\
& \left. + \Delta ( {\bm x}_1 , {\bm b}_1 , {\bm x}_2 , {\bm b}_2 ) +
  \Delta ( {\bm x}_1 , {\bm b}_1 , {\bm b}_2 , {\bm y}_2 ) + \Delta (
  {\bm b}_1 , {\bm y}_1 , {\bm x}_2 , {\bm b}_2 ) + \Delta ( {\bm b}_1
  , {\bm y}_1 , {\bm b}_2 , {\bm y}_2 ) \right].
\end{align}
\eq{eq:2glue_prod_delta} along with the expression for $\Delta$ in
\eq{Delta_exp} are our most complete results for the contribution to
the two-gluon production cross section coming from
\eq{eq:2glue_prod_main} in the quasi-classical regime of the
heavy--light ion collisions in the large-$N_c$ limit. The evaluation
of the full \eq{eq:2glue_prod_delta} appears to be rather involved and
is left for the future work.

Instead we will expand \eq{Delta_exp} and use the result in
\eq{eq:2glue_prod_delta} to obtain the correlated gluon production at
the lowest non-trivial order. This result can be used to elucidate the
structure of the long-range rapidity correlations, along with the
comparison to the existing expressions in the literature.

At the lowest non-trivial order in $D_i$'s (that is, since each $D_i$
represents a two-gluon exchange with the target, at the lowest order
in the number of gluon exchanges corresponding to $k_{1T}, k_{2T} \gg
Q_{s0}$), \eq{Delta_exp} becomes
\begin{align}
  \label{eq:Delta_LO}
  \Delta ( {\bm x}_1 , {\bm x}_2 , {\bm x}_3 , {\bm x}_4 ) \approx
  \frac{(D_3 - D_2)^2}{2 \, N_c^2}.
\end{align}
Substituting this into \eq{eq:2glue_prod_delta} we can further
simplify the expression by assuming that for the connected diagrams
that contribute to the $\Delta$'s one has ${\bm b}_1$ and ${\bm b}_2$
perturbatively close to each other, with the typical separation
between these two impact parameters much smaller than the nucleon
radius. Since the nuclear profile functions for a large nucleus do not
vary much over perturbatively short distances we can put ${\bm b}_1
\approx {\bm b}_2 \approx {\bm b}$ in the arguments of $T_1$'s and
$Q_{s0}$, where ${\bm b} \equiv ({\bm b}_1 + {\bm b}_2)/2$. Defining
$\Delta {\bm b} \equiv {\bm b}_1 - {\bm b}_2$ we can write $d^2 b_1 \,
d^2 b_2 = d^2 b \, d^2 \Delta b$. The integral over $\Delta {\bm b}$
can be then carried out with the help of a Fourier transform
\begin{align}
  \label{eq:Fourier}
  \frac{1}{4} \, |{\bm x}|^2 \, \ln \left( \frac{1}{|{\bm x}| \,
      \Lambda} \right) = \int \frac{d^2 l}{2 \, \pi} \, ( 1 - e^{i \,
    {\bm l} \cdot {\bm x}}) \, \frac{1}{({\bm l}^2)^2}
\end{align}
used to replace $(D_3 - D_2)^2$ in \eq{eq:Delta_LO} and, hence, all
$\Delta$'s in \eq{eq:2glue_prod_delta} as a double Fourier-integral:
for instance
\begin{align}
  \label{eq:Delta_sample}
  \Delta( {\bm x}_1 , {\bm y}_1 , {\bm x}_2 , {\bm y}_2 ) =
  \frac{Q_{s0}^4}{2 \, N_c^2} \int \frac{d^2 l \, d^2 l'}{(2 \pi)^2}
  \, \, \frac{1}{({\bm l}^2)^2} \, \frac{1}{({\bm l}'^2)^2} \, \left[
    e^{i \, {\bm l} \cdot ({\bm x}_1 - {\bm x}_2)} + e^{i \, {\bm l}
      \cdot ({\bm y}_1 - {\bm y}_2)} - e^{i \, {\bm l} \cdot ({\bm
        x}_1 - {\bm y}_2)} - e^{i \, {\bm l} \cdot ({\bm y}_1 - {\bm
        x}_2)} \right] \notag \\
  \times \, \left[ e^{- i \, {\bm l}' \cdot ({\bm x}_1 - {\bm x}_2)} +
    e^{-i \, {\bm l}' \cdot ({\bm y}_1 - {\bm y}_2)} - e^{- i \, {\bm
        l}' \cdot ({\bm x}_1 - {\bm y}_2)} - e^{- i \, {\bm l}' \cdot
      ({\bm y}_1 - {\bm x}_2)} \right] \notag \\
  = \frac{Q_{s0}^4}{2 \, N_c^2} \int \frac{d^2 l \, d^2 l'}{(2 \pi)^2}
  \, \, \frac{1}{({\bm l}^2)^2} \, \frac{1}{({\bm l}'^2)^2} \, \left[
    e^{i \, {\bm l} \cdot ({\tilde {\bm x}}_1 - {\tilde {\bm x}}_2)} +
    e^{i \, {\bm l} \cdot ({\tilde {\bm y}}_1 - {\tilde {\bm y}}_2)} -
    e^{i \, {\bm l} \cdot ({\tilde {\bm x}}_1 - {\tilde {\bm y}}_2)} -
    e^{i \, {\bm l} \cdot ({\tilde {\bm y}}_1 - {\tilde {\bm
          x}_2)}} \right] \notag \\
  \times \, \left[ e^{- i \, {\bm l}' \cdot ({\tilde {\bm x}}_1 -
      {\tilde {\bm x}}_2)} + e^{-i \, {\bm l}' \cdot ({\tilde {\bm
          y}}_1 - {\tilde {\bm y}}_2)} - e^{- i \, {\bm l}' \cdot
      ({\tilde {\bm x}}_1 - {\tilde {\bm y}}_2)} - e^{- i \, {\bm l}'
      \cdot ({\tilde {\bm y}}_1 - {\tilde {\bm x}}_2)} \right] \, e^{i
    \, ({\bm l} - {\bm l}') \cdot \Delta {\bm b}}
\end{align}
where we employed the substitution 
\begin{align}
  \label{eq:change}
  {\bm {\tilde x}}_1 = {\bm x}_1 - {\bm b}_1, \ \ {\bm {\tilde y}}_1 =
  {\bm y}_1 - {\bm b}_1, \ \ {\bm {\tilde x}}_2 = {\bm x}_2 - {\bm
    b}_2, \ \ {\bm {\tilde y}}_2 = {\bm y}_2 - {\bm b}_1.
\end{align}
Performing similar substitutions for all $\Delta$'s in
\eq{eq:2glue_prod_delta}, we can integrate over $\Delta {\bm b}$,
${\tilde {\bm x}}_1$, ${\tilde {\bm x}}_2$, ${\tilde {\bm y}}_1$,
${\tilde {\bm y}}_2$, and ${\bm l}'$. After some algebra one arrives
at (cf. \cite{Dusling:2009ni,Dumitru:2010iy})
\begin{align} 
\label{eq:2glue_prod_LO} 
\frac{d \sigma_{square}^{(corr)}}{d^2 k_1 dy_1 d^2 k_2 dy_2}\bigg|_{LO} =
\frac{\as^2}{4 \, \pi^4} \, \int d^2 B \, d^2 b \, [T_1 ({\bm B} -
{\bm b})]^2 \, \frac{Q_{s0}^4 ({\bm b})}{{\bm k}_1^2 \, {\bm k}_2^2}
\, \int\limits_\Lambda \frac{d^2 l}{({\bm l}^2)^2} \, \left[
  \frac{1}{({\bm k}_1 - {\bm l})^2 \, ({\bm k}_2 + {\bm l})^2} +
  \frac{1}{({\bm k}_1 - {\bm l})^2 \, ({\bm k}_2 - {\bm l})^2}
\right],
\end{align}
where the subscript $LO$ denotes the lowest-order cross section.

\begin{figure}[hb]
  \includegraphics[width= 0.9 \textwidth]{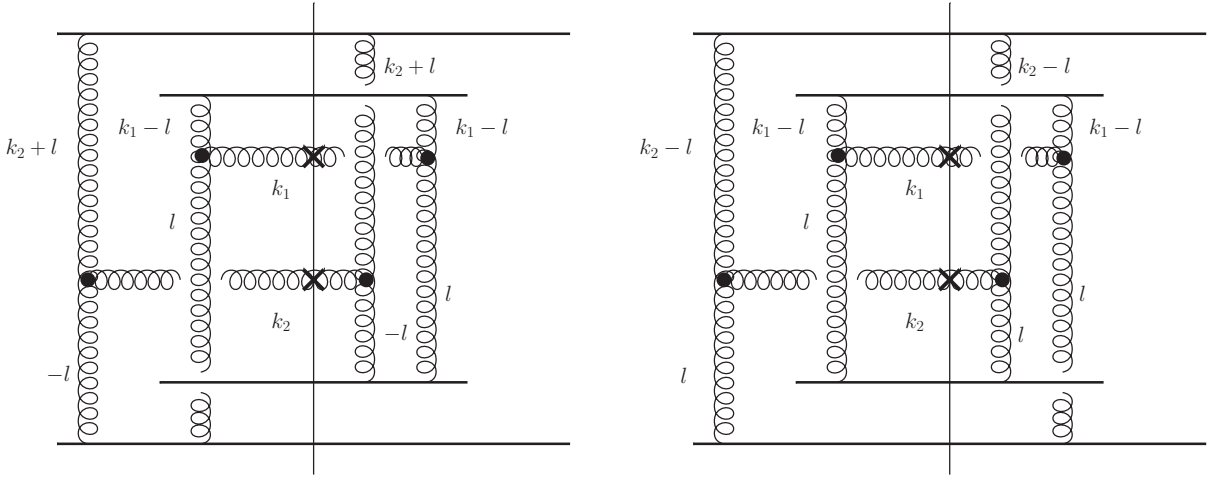}
  \caption{Examples of diagrams generating contributions to
    \eq{eq:2glue_prod_LO}: the left panel represent the away-side
    correlations (the first term in \eq{eq:2glue_prod_LO}), while the
    right panel contributes near-side correlations (the second term in
    \eq{eq:2glue_prod_LO}). The $t$-channel gluon momenta flow toward
    the triple-gluon vertices to the left of the cut, and away from
    those vertices to the right of the cut.}
\label{glgraph1} 
\end{figure}

Equation \eqref{eq:2glue_prod_LO} is illustrated in \fig{glgraph1} by
regular Feynman diagrams that contribute to its right-hand-side
(cf. \cite{Dusling:2009ni,Dumitru:2010iy}): these diagrams are
referred to as the ``glasma'' graphs in the literature. The momenta of
the gluon lines are labeled in \fig{glgraph1}, and the triple gluon
vertices, marked by the dark circles, are the effective Lipatov
vertices. These particular graphs contribute to the first and the
second terms in the square brackets of \eq{eq:2glue_prod_LO}
correspondingly, and can be calculated by taking two Lipatov vertices
squared. 

The obtained expression \eqref{eq:2glue_prod_LO} contains both the
near-side and away-side azimuthal correlations
\cite{Dusling:2009ni,Dumitru:2010iy}: clearly the first term in the
square brackets of \eq{eq:2glue_prod_LO} contains poles at ${\bm l} =
{\bm k}_1$ and ${\bm l} = - {\bm k}_2$, which, after integration over
$l$, lead to a contribution\footnote{Note that both terms in
  \eq{eq:2glue_prod_LO} also contain a pole at ${\bm l} =0$, which
  leads to a correlated contribution independent of the azimuthal
  angle between ${\bm k}_1$ and ${\bm k}_2$.}
\begin{align}
  \label{eq:b2b}
  \sim \frac{1}{({\bm k}_1 + {\bm k}_2)^2},
\end{align}
characteristic of the away-side correlations.

The second term in the square brackets of \eq{eq:2glue_prod_LO} has
poles at ${\bm l} = {\bm k}_1$ and ${\bm l} = {\bm k}_2$, yielding a
contribution
\begin{align}
  \label{eq:near_side}
  \sim \frac{1}{({\bm k}_1 - {\bm k}_2)^2},
\end{align}
indicating near-side correlations. Note that all correlations are
long-range in rapidity since the cross section
\eqref{eq:2glue_prod_LO} is rapidity-independent.

Now we turn our attention to \eq{crossed_xsect}. There the cross
section contribution itself is $1/N_c^2$-suppressed as compared to the
leading (uncorrelated) part of \eq{eq:2glue_prod_main}: hence we need
to evaluate the interaction with the target in \eqref{crossed_xsect}
using the large-$N_c$ limit. We work in the same quasi-classical MV/GM
approximation, along with the large-$N_c$ limit. The fundamental
(quark) quadrupole amplitude was evaluated in this approximation in
\cite{JalilianMarian:2004da} (see Eq. (14) there), yielding
\begin{align}
  \label{eq:quad_quark}
  Q_{quark} ({\bm x}_1, {\bm x}_2, {\bm x}_3, {\bm x}_4) = e^{D_1/2} +
  \frac{D_3 - D_2}{D_1 - D_3} \, \left[ e^{D_1/2} - e^{D_3/2} \right].
\end{align}
Since the adjoint (gluon) quadrupole \eqref{quad_def} in the
large-$N_c$ limit is simply
\begin{align}
  \label{eq:quad_rel}
  Q ({\bm x}_1, {\bm x}_2, {\bm x}_3, {\bm x}_4) = \left[ Q_{quark}
    ({\bm x}_1, {\bm x}_2, {\bm x}_3, {\bm x}_4) \right]^2
\end{align}
we get
\begin{align}
  \label{eq:quad_MV}
  Q ({\bm x}_1, {\bm x}_2, {\bm x}_3, {\bm x}_4) = \left[e^{D_1/2} +
    \frac{D_3 - D_2}{D_1 - D_3} \, \left( e^{D_1/2} - e^{D_3/2}
    \right) \right]^2.
\end{align}
Equations \eqref{eq:quad_MV} and \eqref{eq:SG_GM}, when used in
\eq{crossed_xsect}, give us the remaining contribution to the
two-gluon production cross section in the quasi-classical
approximation, which has to be added to \eq{eq:2glue_prod_delta}.

Again the full expression \eqref{crossed_xsect} is hard to evaluate:
instead, similar to \eq{eq:2glue_prod_LO}, we will consider the limit
of large $k_{1T} = |{\bm k}_1|$ and $k_{2T} = |{\bm k}_2|$: $k_1, k_2
\gg Q_{s0}$. Expanding \eq{eq:quad_MV} in the powers of $D_i$'s yields
\begin{align}
  \label{eq:quad_exp}
  Q ({\bm x}_1, {\bm x}_2, {\bm x}_3, {\bm x}_4) = 1 + D_1 - D_2 + D_3
  + \frac{1}{4} \, \left[ 2 \, D_1^2 + D_2^2 + 2 \, D_3^2 - 3 \, D_1
    \, D_2 - 3 \, D_2 \, D_3 + 3 \, D_1 \, D_3 \right] + O \left(
    D_i^3 \right).
\end{align}
Using this result along with a similar expansion for \eq{eq:SG_GM} in
\eq{crossedinteraction}, and employing again the Fourier transform
\eqref{eq:Fourier} along with the substitution \eqref{eq:change} one
can write
\begin{align}
  \label{eq:intQ}
  Int_{crossed} ( {\bm x}_1, {\bm y}_1 , {\bm b}_1 , {\bm x}_2 , {\bm
    y}_2, {\bm b}_2) = Q_{s0}^4 \, \int \frac{d^2 l \, d^2 l'}{(2
    \pi)^2} \, \, \frac{1}{({\bm l}^2)^2} \, \frac{1}{({\bm l}'^2)^2}
  \, \bigg\{ e^{i \, ({\bm l} - {\bm l}') \cdot \Delta {\bm b}} \,
  \left( 1 - e^{i \, {\bm l}' \cdot {\bm {\tilde x}}_2} \right) \,
  \left( 1 - e^{- i \, {\bm l} \cdot {\bm {\tilde y}}_2} \right)
  \notag \\ \times \, \bigg[ \frac{1}{2} \, \left( 1 - e^{- i \, {\bm
        l}' \cdot {\bm {\tilde x}}_1} \right) \, \left( 1 - e^{i \,
      {\bm l} \cdot {\bm {\tilde y}}_1} \right) + \left( 1 - e^{i \,
      {\bm l} \cdot {\bm {\tilde x}}_1} \right) \, \left( 1 - e^{- i
      \, {\bm l}' \cdot {\bm {\tilde y}}_1} \right) \bigg] \notag \\ +
  \left( 1 - e^{i \, {\bm l} \cdot {\bm {\tilde x}}_1} \right) \,
  \left( 1 - e^{- i \, {\bm l}' \cdot {\bm {\tilde x}}_2} \right) \,
  \left( 1 - e^{- i \, {\bm l} \cdot {\bm {\tilde y}}_1} \right) \,
  \left( 1 - e^{i \, {\bm l}' \cdot {\bm {\tilde y}}_2} \right)
  \bigg\}
\end{align}
where the ${\bm l} \leftrightarrow {\bm l}'$, ${\bm l} \leftrightarrow
- {\bm l}$, and ${\bm l}' \leftrightarrow - {\bm l}'$ symmetries of
the integrand were utilized to cast the expression in its present
form. Using \eq{eq:intQ} to replace the interaction with the target in
\eq{crossed_xsect} and integrating over ${\bm {\tilde x}}_1$, ${\bm
  {\tilde x}}_2$, ${\bm {\tilde y}}_1$, ${\bm {\tilde y}}_2$, and, in
some terms, over ${\bm l}'$, leads to the following result
\begin{align}
  \label{eq:rap_corr2}
  \frac{d \sigma_{crossed}}{d^2 k_1 dy_1 d^2 k_2 dy_2}\bigg|_{LO} =
  \frac{d \sigma_{crossed}^{(corr)}}{d^2 k_1 dy_1 d^2 k_2
    dy_2}\bigg|_{LO} + \frac{d \sigma_{HBT}}{d^2 k_1 dy_1 d^2 k_2
    dy_2}\bigg|_{LO}
\end{align}
where
\begin{align}\label{eq:rap_corr3}
  \frac{d \sigma^{(corr)}_{crossed}}{d^2 k_1 dy_1 d^2 k_2
    dy_2}\bigg|_{LO} = & \frac{\as^2}{32 \, \pi^4} \, \int d^2 B \,
  d^2 b \, [T_1 ({\bm B} - {\bm b})]^2 \, \frac{Q_{s0}^4 ({\bm
      b})}{{\bm k}_1^2 \, {\bm k}_2^2} \, \int \frac{d^2 l}{({\bm
      l}^2)^2 \, (({\bm l} - {\bm k}_1 + {\bm k}_2)^2)^2 \, (({\bm
      k}_1 - {\bm l})^2)^2 \, (({\bm k}_2 + {\bm l})^2)^2} \notag \\ &
  \times \big\{ \left[ {\bm l}^2 \, ({\bm k}_2 + {\bm l})^2 + ({\bm
      k}_1 - {\bm l})^2 \, ({\bm l} - {\bm k}_1 + {\bm k}_2)^2 - {\bm
      k}_1^2 \, ({\bm k}_2 - {\bm k}_1 + 2 \, {\bm l})^2 \right]
  \notag \\ & \times \, \left[ {\bm l}^2 \, ({\bm k}_1 - {\bm l})^2 +
    ({\bm k}_2 + {\bm l})^2 \, ({\bm l} - {\bm k}_1 + {\bm k}_2)^2 -
    {\bm k}_2^2 \, ({\bm k}_2 - {\bm k}_1 + 2 \, {\bm l})^2 \right]
  \notag \\ & + 4 \, {\bm l}^2 \, ({\bm l} - {\bm k}_1 + {\bm k}_2)^2
  \, \left[ (({\bm k}_1 - {\bm l})^2)^2 + (({\bm k}_2 + {\bm l})^2)^2
  \right] \big\} + ({\bm k}_2 \rightarrow - {\bm k}_2)
\end{align}
and
\begin{align} 
\label{eq:HBT_LO} 
& \frac{d \sigma_{HBT}}{d^2 k_1 dy_1 d^2 k_2 dy_2}\bigg|_{LO} =
\frac{\as^2}{16 \, \pi^4} \, \int d^2 B \, d^2 b \, [T_1 ({\bm B} -
{\bm b})]^2 \, \frac{Q_{s0}^4 ({\bm b})}{{\bm k}_1^2 \, {\bm k}_2^2}
\, \left[ \delta^2 ({\bm k}_1 - {\bm k}_2) + \delta^2 ({\bm k}_1 +
  {\bm k}_2) \right] \, \notag \\ & \times \, \int\limits_\Lambda
\frac{d^2 l \, d^2 l'}{({\bm l}^2)^2 \, ({\bm l '}^2)^2 \, (({\bm k}_1
  - {\bm l})^2)^2 \, (({\bm k}_1 + {\bm l'})^2)^2} \, \left[ {\bm l}^2
  \, ({\bm k}_1 + {\bm l'})^2 + {\bm l'}^2 \, ({\bm k}_1 - {\bm l})^2
  - {\bm k}_1^2 \, ({\bm l} + {\bm l'})^2 \right]^2.
\end{align}
We will explain the origin of the ``HBT'' label on the cross section
in \eq{eq:HBT_LO} in a little while below: we defer the analysis of
\eq{eq:HBT_LO} until then.

First we concentrate on the expression \eqref{eq:rap_corr3}. We note
that by shifting the integration momentum $\bm l$ one could reduce the
second term in the curly brackets (together with the ${\bm k}_2
\rightarrow - {\bm k}_2$ term) to that in \eq{eq:2glue_prod_LO}, thus
doubling \eq{eq:2glue_prod_LO}. That term is still described by the
diagrams of the type shown in \fig{glgraph1} and gives the near- and
away-side correlations in Eqs. \eqref{eq:b2b} and
\eqref{eq:near_side}.

\begin{figure}[ht]
  \includegraphics[width= 0.45 \textwidth]{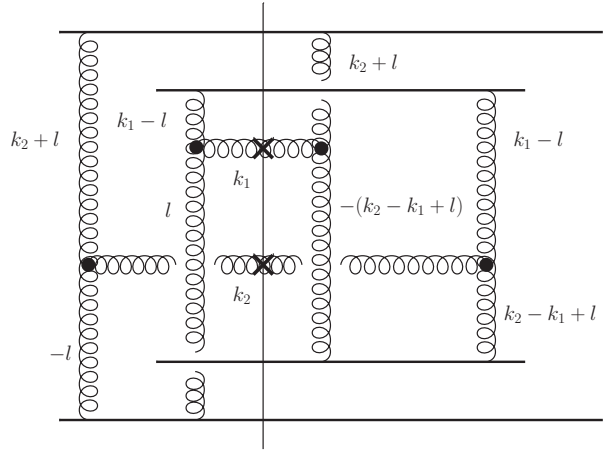}
  \caption{An example of a diagram giving a contribution to the first
    term in the curly brackets of \eq{eq:rap_corr3}.}
\label{glgraph2} 
\end{figure}

The first term in the curly brackets of \eq{eq:rap_corr3} corresponds
to a different class of diagrams, one of which is shown in
\fig{glgraph2}. One can see that the diagram in \fig{glgraph2}
contributes two non-forward ``squares'' of the effective Lipatov
vertices. An analysis of the poles of the integral in the first term
of \eq{eq:rap_corr3} shows that it contains similar near- and
away-side correlations to that in Eqs. \eqref{eq:b2b} and
\eqref{eq:near_side}, though with an additional enhancement due to a
prefactor, 
\begin{align}
  \label{eq:near_side2}
  \sim [2 ({\bm k}_1 \cdot {\bm k}_2)^2 - {\bm k}_1^2 \, {\bm k}_2^2]
  \, \left[\frac{1}{({\bm k}_1 - {\bm k}_2)^2} + \frac{1}{({\bm k}_1 +
      {\bm k}_2)^2} \right].
\end{align}
The origin of this contribution to correlations is in the non-forward
squares of the Lipatov vertices, which is similar to the mechanism for
generating long-range rapidity correlations proposed in
\cite{Levin:2011fb} using two BFKL ladders with non-zero momentum
transfer. Note, however, that we do not obtain the correlation
proportional to the first power of ${\bm k}_1 \cdot {\bm k}_2$
advocated in \cite{Levin:2011fb}, perhaps due to the lowest-order
nature of the result \eqref{eq:rap_corr3}. It appears that the
correlations \eqref{eq:near_side2} resulting from the lowest-order
diagrams like the one depicted in \fig{glgraph2} have not been
considered in the literature yet.

Note that the away- and the near-side correlations enter
Eqs. \eqref{eq:2glue_prod_LO} and \eqref{eq:rap_corr3} on equal
footing: in fact, one could be obtained from another by a simple ${\bm
  k}_2 \rightarrow - {\bm k}_2$ substitution in either of those
expressions. While evaluating the integrals over $l$ in
Eqs. \eqref{eq:2glue_prod_LO} and \eqref{eq:rap_corr3} analytically
appears to be rather algebra-intensive, we do not need to do this to
observe that, once the $l$-integral is carried out for one of the
terms in the square brackets of \eq{eq:2glue_prod_LO}, the answer for
the other term is obtained by substituting ${\bm k}_2 \rightarrow -
{\bm k}_2$ in the result. Equation \eqref{eq:rap_corr3} simply
contains an additive ${\bm k}_2 \rightarrow - {\bm k}_2$ term.

Since the correlated cross sections \eqref{eq:2glue_prod_LO} and
\eqref{eq:rap_corr3} are sums of two terms related by the ${\bm k}_2
\rightarrow - {\bm k}_2$ substitution and are symmetric under the
${\bm k}_1 \leftrightarrow {\bm k}_2$ interchange, we conclude that
they are functions only of even powers of ${\bm k}_1$ and ${\bm k}_2$,
that is functions of ${\bm k}_1^2$, ${\bm k}_2^2$, $({\bm k}_1 \cdot
{\bm k}_2)^2$, and possibly $({\bm k}_1 \times {\bm k}_2)^2$. Clearly
this implies that the Fourier series representation of
Eqs. \eqref{eq:2glue_prod_LO} and \eqref{eq:rap_corr3} would only
contain even cosine harmonics of the azimuthal angle, that is
\begin{align}
  \label{eq:series}
  \frac{d \sigma^{(corr)}}{d^2 k_1 dy_1 d^2 k_2 dy_2}\bigg|_{LO}
  \equiv \frac{d \sigma^{(corr)}_{square}}{d^2 k_1 dy_1 d^2 k_2
    dy_2}\bigg|_{LO} + \frac{d \sigma^{(corr)}_{crossed}}{d^2 k_1 dy_1
    d^2 k_2 dy_2}\bigg|_{LO} \sim \sum_{n=0}^\infty c_n (k_{1T},
  k_{2T}) \, \cos (2 \, n \, \Delta \phi)
\end{align}
where $\Delta \phi = \phi_1 - \phi_2$ is the angle between momenta
${\bm k}_1$ and ${\bm k}_2$ while $k_{1T} = |{\bm k}_1|$ and $k_{2T} =
|{\bm k}_2|$. Here $c_n (k_1, k_2)$ are some coefficient to be
determined by an exact calculation. It is quite interesting that only
even harmonics contribute to the correlated cross section in
\eq{eq:series}. Let us stress here that we have not made any
assumptions about the centrality of the collision: we do not have an
almond-shaped overlap of the two nuclei. In fact we integrate over all
impact parameters $\bm B$. (Also the impact parameter dependence
factorizes from the rest of the expression.) Hence the correlation in
\eq{eq:series} is not caused by the geometry of the collision.

To construct the correlation function one has to use
Eqs. \eqref{eq:2glue_prod_LO} and \eqref{eq:rap_corr3} in
\eq{corr_main}. In the latter, the uncorrelated two-gluon production
would dominate in the denominator of the normalization factor, that
is, in the denominator of the first factor on its right-hand side. The
single-gluon production cross-section \eqref{crosssection12} at the
lowest order is equal to (see \cite{Jalilian-Marian:2005jf} and
references therein)
\begin{align}
  \label{eq:1GLO}
  \left\langle \frac{d \sigma^{pA_2}}{d^2 k \, dy \, d^2 b}
  \right\rangle = \frac{\as \, C_F}{\pi^2} \, \frac{Q_{s0}^2 ({\bm
      b})}{k_T^4} \, \ln \frac{k_T^2}{\Lambda^2}.
\end{align}
Using this along with the sum of Eqs. \eqref{eq:2glue_prod_LO} and
\eqref{eq:rap_corr3} in \eq{corr_main} and taking the large-$N_c$
limit in \eq{eq:1GLO} we get
\begin{align}
  \label{eq:corr_LO}
  & C ({\bm k}_1, y_1, {\bm k}_2, y_2)\big|_{LO} = \frac{1}{N_c^2} \,
  \frac{\int d^2 B \, d^2 b \, [T_1 ({\bm B} - {\bm b})]^2 \, Q_{s0}^4
    ({\bm b})}{\int d^2 B \, d^2 b_1 \, d^2 b_2 \, T_1 ({\bm B} - {\bm
      b}_1) \, T_1 ({\bm B} - {\bm b}_2) \, Q_{s0}^2 ({\bm b}_1) \,
    Q_{s0}^2 ({\bm b}_2) } \notag \\ & \times \, \frac{{\bm k}_1^2 \,
    {\bm k}_2^2}{\ln \frac{k_1^2}{\Lambda^2} \, \ln
    \frac{k_2^2}{\Lambda^2}} \, \bigg\{ 2 \, \int\limits_\Lambda
  \frac{d^2 l}{({\bm l}^2)^2} \, \left[ \frac{1}{({\bm k}_1 - {\bm
        l})^2 \, ({\bm k}_2 + {\bm l})^2} + \frac{1}{({\bm k}_1 - {\bm
        l})^2 \, ({\bm k}_2 - {\bm l})^2} \right] \notag \\ & +
  \frac{1}{8} \, \bigg[ \int \frac{d^2 l}{({\bm l}^2)^2 \, (({\bm l} -
    {\bm k}_1 + {\bm k}_2)^2)^2 \, (({\bm k}_1 - {\bm l})^2)^2 \,
    (({\bm k}_2 + {\bm l})^2)^2} \, \left[ {\bm l}^2 \, ({\bm k}_2 +
    {\bm l})^2 + ({\bm k}_1 - {\bm l})^2 \, ({\bm l} - {\bm k}_1 +
    {\bm k}_2)^2 - {\bm k}_1^2 \, ({\bm k}_2 - {\bm k}_1 + 2 \, {\bm
      l})^2 \right] \notag \\ & \times \, \left[ {\bm l}^2 \, ({\bm
      k}_1 - {\bm l})^2 + ({\bm k}_2 + {\bm l})^2 \, ({\bm l} - {\bm
      k}_1 + {\bm k}_2)^2 - {\bm k}_2^2 \, ({\bm k}_2 - {\bm k}_1 + 2
    \, {\bm l})^2 \right] + ({\bm k}_2 \rightarrow - {\bm k}_2) \bigg]
  \bigg\}.
\end{align}

To evaluate this further let us assume for a moment that both
colliding nuclei are simply cylinders oriented along the collision
axis, such that the nuclear profile functions of the nuclei are $T_i
({\bm b}) = 2 \, \rho \, R_i \, \theta (R_i - b)$ where $i = 1,2$
labels the nuclei, $\rho$ is the (constant) nucleon number density in
the nucleus, $R_1$ and $R_2$ are the radii of the projectile and
target nuclei, and the cylindrical nucleus is assumed to have length
$2 R_i$ along its axis. Assuming that both nuclei are large and
neglecting the edge effects, we can neglect the ${\bm b}_1$ and ${\bm
  b}_2$ dependence in these nuclear profile functions of both
nuclei. Since the gluon saturation scale in the MV model is $Q_{s0}^2
= 4 \, \pi \, \as^2 \, T_2 ({\bm b})$ we obtain in the $R_1 \ll R_2$
limit
\begin{align}
  \label{eq:corr_LO_cyl}
  & C ({\bm k}_1, y_1, {\bm k}_2, y_2)\big|_{LO} = \frac{1}{N_c^2 \,
    \pi R_1^2} \, \frac{{\bm k}_1^2 \, {\bm k}_2^2}{\ln
    \frac{k_1^2}{\Lambda^2} \, \ln \frac{k_2^2}{\Lambda^2}} \, \bigg\{
  2 \, \int\limits_\Lambda \frac{d^2 l}{({\bm l}^2)^2} \, \left[
    \frac{1}{({\bm k}_1 - {\bm l})^2 \, ({\bm k}_2 + {\bm l})^2} +
    \frac{1}{({\bm k}_1 - {\bm l})^2 \, ({\bm k}_2 - {\bm l})^2}
  \right] \notag \\ & + \frac{1}{8} \, \bigg[ \int \frac{d^2 l}{({\bm
      l}^2)^2 \, (({\bm l} - {\bm k}_1 + {\bm k}_2)^2)^2 \, (({\bm
      k}_1 - {\bm l})^2)^2 \, (({\bm k}_2 + {\bm l})^2)^2} \, \left[
    {\bm l}^2 \, ({\bm k}_2 + {\bm l})^2 + ({\bm k}_1 - {\bm l})^2 \,
    ({\bm l} - {\bm k}_1 + {\bm k}_2)^2 - {\bm k}_1^2 \, ({\bm k}_2 -
    {\bm k}_1 + 2 \, {\bm l})^2 \right] \notag \\ & \times \, \left[
    {\bm l}^2 \, ({\bm k}_1 - {\bm l})^2 + ({\bm k}_2 + {\bm l})^2 \,
    ({\bm l} - {\bm k}_1 + {\bm k}_2)^2 - {\bm k}_2^2 \, ({\bm k}_2 -
    {\bm k}_1 + 2 \, {\bm l})^2 \right] + ({\bm k}_2 \rightarrow -
  {\bm k}_2) \bigg] \bigg\}.
\end{align}
Indeed the correlator is suppressed by a power of $N_c^2$ and a power
of the cross sectional area of the projectile nucleus $\pi R_1^2$, as
discussed in the literature
\cite{Dusling:2009ni,Dumitru:2010iy,Dumitru:2010mv,Kovner:2010xk,KovnerLublinsky12}.

While our conclusion about the correlator \eqref{eq:corr_LO_cyl} and
the cross section \eq{eq:series} contributing only to even Fourier
harmonics has been verified above at the lowest order only, it is true
for the full two-gluon production cross section in heavy-light ion
collisions \eqref{eq_all}. This can be seen by noticing that ${\bm
  k}_2 \rightarrow - {\bm k}_2$ substitution in
\eq{eq:2glue_prod_delta} does not change the cross section, as it is
equivalent to the ${\bm x}_2 \leftrightarrow {\bm y}_2$ interchange of
the integration variables. The integrand of \eq{eq:2glue_prod_delta}
(or, equivalently, of \eq{eq:2glue_prod_main}) is invariant under
${\bm x}_2 \leftrightarrow {\bm y}_2$ interchange since the gluon is
its own anti-particle such that $Tr \left[ U_{\bm x} \, U^\dagger_{\bm
    y} \right] = Tr \left[ U_{\bm y} \, U^\dagger_{\bm x}
\right]$. The expression in \eq{eq:rap_corr3} is explicitly invariant
under ${\bm k}_2 \rightarrow - {\bm k}_2$ substitution. Hence the net
correlated cross section \eqref{eq_all} is an even function of ${\bm
  k}_2$. Note also that Eqs. \eqref{eq:2glue_prod_main} and
\eqref{crossed_xsect} are ${\bm k}_1 \leftrightarrow {\bm
  k}_2$-symmetric: in \eq{eq:2glue_prod_main} and in the term arising
from the first exponential in \eq{crossed_xsect} the symmetry is a
consequence of the symmetry of the integrand under the simultaneous
${\bm x}_1 \leftrightarrow {\bm x}_2$, ${\bm y}_1 \leftrightarrow {\bm
  y}_2$, and ${\bm b}_1 \leftrightarrow {\bm b}_2$ interchanges. The
term multiplying the second exponential in \eq{crossed_xsect} is ${\bm
  k}_1 \leftrightarrow {\bm k}_2$-symmetric due to the ${\bm x}_1
\leftrightarrow {\bm y}_1$, ${\bm x}_2 \leftrightarrow {\bm y}_2$
symmetry of the integrand, which follows from the following property
of the quadrupole operator: $Tr \left[ U_{{\bm x}_1} \,
  U^\dagger_{{\bm y}_1} \, U_{{\bm x}_2} \, U^\dagger_{{\bm y}_2}
\right] = Tr \left[ U_{{\bm y}_2} \, U^\dagger_{{\bm x}_2} \, U_{{\bm
      y}_1} \, U^\dagger_{{\bm x}_1} \right]$. We see that the net
cross section \eqref{eq_all} is decomposable into a Fourier series
with even harmonics only. Therefore, the correlation function in the
heavy-light ion collision can be also written as an even-harmonics
series
\begin{align}
  \label{eq:Cseries}
  C ({\bm k}_1, y_1, {\bm k}_2, y_2)\big|_{A_2 \gg A_1 \gg 1} \sim
  \sum_{n=0}^\infty d_n (k_{1T}, k_{2T}) \, \cos (2 \, n \, \Delta
  \phi)
\end{align}
with some coefficients $d_n$. This conclusion also seems to hold in
the case of classical gluon fields produced in a collision of two
heavy ions, as can be seen from the result of the full numerical
simulation of two-gluon production in heavy ion collisions due to
classical gluon field carried out in \cite{Lappi:2009xa}. The
correlators in Figs.~9 and 10 of \cite{Lappi:2009xa} do appear to have
similar-looking maxima at $\Delta \phi =0$ and $\Delta \phi =\pi$
(within the accuracy of the numerical error bars), though a more
careful analysis is needed to figure out if our conclusion is true for
two colliding heavy ions.

\begin{figure}[h]
  \includegraphics[width= 0.45 \textwidth]{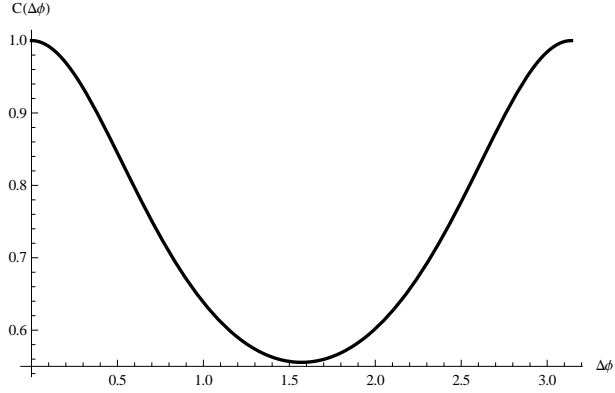}
  \caption{A toy azimuthal two-gluon correlation function motivated by
    the calculation in this Section. Vertical scale is arbitrary.}
\label{corr_toy} 
\end{figure}

To better visualize the correlator let us envision a toy model in
which multiple rescatterings (and other saturation effects) regulate
the singularities at ${\bm k}_1 = \pm {\bm k}_2$ by the saturation
scale $Q_{s0}$ in such a way that the correlator can be modeled as
proportional to
\begin{align}
  \label{eq:corr_model}
  C_{\mbox{toy} \ \mbox{model}} ({\bm k}_1, y_1, {\bm k}_2, y_2) \sim
  \frac{1}{({\bm k}_1 - {\bm k}_2)^2 + Q_{s0}^2} + \frac{1}{({\bm k}_1
    + {\bm k}_2)^2 + Q_{s0}^2}.
\end{align}
This correlator is plotted in \fig{corr_toy} using arbitrary units
along the vertical axis as a function of the azimuthal angle $\Delta
\phi$ between ${\bm k}_1$ and ${\bm k}_2$ for $k_1 = k_2 =
Q_{s0}$. The shape illustrates what the full correlation function may
look like, having identical near- and away-side correlation peaks.

Our conclusion in this Section is that the saturation/CGC dynamics in
nuclear collisions appears to generate the long-range rapidity
correlations which have identical maxima at both $\Delta \phi =0$ and
$\Delta \phi =\pi$ in the azimuthal angle. Such correlations are
non-flow in nature, since they do not arise due to almond-shaped
geometry of the collisions. (In fact the correlations should persist
for the most central heavy ion collisions, though they should be
suppressed by an inverse power of the overlap area as discussed above
and in
\cite{Dusling:2009ni,Dumitru:2010iy,Dumitru:2010mv,Kovner:2010xk,KovnerLublinsky12}.)
However, since the elliptic flow observable $v_2$ (even in the
reaction-plane method) is determined from two-particle correlations,
it is possible that the correlations discussed here may contribute to
the elliptic flow (and to higher-order even-harmonics flow observables
$v_{2n}$) measured experimentally. It is, therefore, very important to
experimentally separate our initial-state correlations from the
late-time QGP effects.

Naturally the cumulant analysis \cite{Borghini:2000sa,Borghini:2001vi}
is likely to remove these two-particle non-flow correlations from the
flow observables. However, the effectiveness of the cumulant method
may again depend on the collision geometry. Let us illustrate this by
considering the fourth order cumulant for elliptic flow, defined by
\cite{Borghini:2000sa,Borghini:2001vi}
\begin{align}
  \label{eq:4cumulant}
  c_2 \{ 4 \} \equiv \left\langle e^{2 \, i \, (\phi_1 + \phi_2 -
      \phi_3 - \phi_4)} \right\rangle - \left\langle e^{2 \, i \,
      (\phi_1 - \phi_3)} \right\rangle \, \left\langle e^{2 \, i \,
      (\phi_2 - \phi_4)} \right\rangle - \left\langle e^{2 \, i \,
      (\phi_1 - \phi_4)} \right\rangle \, \left\langle e^{2 \, i \,
      (\phi_2 - \phi_3)} \right\rangle
\end{align}
where the angle brackets denote event averages along with the
averaging over all angles $\phi_1, \phi_2, \phi_3$, and $\phi_4$ of
the four particles employed in the definition. Using the lowest-order
correlations employed in arriving at \eq{eq:corr_LO} one can
straightforwardly show that the cumulant due to these correlations
only is proportional to
\begin{align}
  \label{eq:LOcumulant}
  c_2 \{ 4 \} \big|_{LO} \propto & \frac{\int d^2 B \, d^2 b_1 \, d^2
    b_2 \, [T_1 ({\bm B} - {\bm b}_1)]^2 \, [T_1 ({\bm B} - {\bm
      b}_2)]^2 \, Q_{s0}^4 ({\bm b}_1) \, Q_{s0}^4 ({\bm b}_2)}{\int
    d^2 B \, d^2 b_1 \, d^2 b_2 \, d^2 b_3 \, d^2 b_4 \, T_1 ({\bm B}
    - {\bm b}_1) \, T_1 ({\bm B} - {\bm b}_2) \, T_1 ({\bm B} - {\bm
      b}_3) \, T_1 ({\bm B} - {\bm b}_4) \, Q_{s0}^2 ({\bm b}_1) \,
    Q_{s0}^2 ({\bm b}_2) \, Q_{s0}^2 ({\bm b}_3) \, Q_{s0}^2 ({\bm
      b}_4)} \notag \\
  & - \left[ \frac{\int d^2 B \, d^2 b \, [T_1 ({\bm B} - {\bm b})]^2
      \, Q_{s0}^4 ({\bm b})}{\int d^2 B \, d^2 b_1 \, d^2 b_2 \, T_1
      ({\bm B} - {\bm b}_1) \, T_1 ({\bm B} - {\bm b}_2) \, Q_{s0}^2
      ({\bm b}_1) \, Q_{s0}^2 ({\bm b}_2) } \right]^2
\end{align}
if the 4-particle correlator in \eq{eq:4cumulant} (the correlated part
of the first term on its right-hand-side) is due to the pairwise
correlations of particles $1, 3$ and $2, 4$ or $1, 4$ and
$2,3$. Clearly, in the general case, $c_2 \{ 4 \} \big|_{LO}$ from
\eq{eq:LOcumulant} is non-zero.\footnote{Note, however, that for the
  heavy-light nuclear collision case considered here, $A_1 \ll A_2$,
  and for cylindrical nuclei, one gets $c_2 \{ 4 \} \big|_{LO} =0$.}
Moreover, the fourth order cumulant would contain the azimuthal angles
dependence resulting from \eq{eq:corr_LO} but not shown explicitly in
\eq{eq:LOcumulant}. We see that the geometric correlations from
Sec.~\ref{sec:disc} may prevent complete removal of these non-flow
correlations in the cumulants. Just like with the correlator of
\eq{eq:CfixedB}, the above non-flow correlations can be completely
removed from the cumulants for the fixed impact parameter $\bm B$: if
we fix $\bm B$ in \eq{eq:LOcumulant} (that is, remove all the $d^2 B$
integrations from it), we get $c_2 \{ 4 \} \big|_{LO} =0$, which is
exactly what the cumulant is designed to do
\cite{Borghini:2000sa,Borghini:2001vi} --- completely cancel for
non-flow correlations. Note that since, in the actual experimental
analyses one effectively integrates over $\bm B$ in a given centrality
bin, it is possible that some non-flow correlations \eqref{eq:corr_LO}
would remain in the cumulant \eqref{eq:LOcumulant}. Even fixing $|{\bm
  B}|$ precisely and integrating over the angles of $\bm B$ may
generate a non-zero contribution of these correlations to the
cumulant. The question of the interplay of the true QGP flow and the
non-flow correlations discussed here has to be resolved by a more
detailed numerical study.


To clarify the physical meaning of the cross section obtained in
\eq{eq:HBT_LO} we again consider a collisions of cylindrical
nuclei. With this simplification we can consider one of the terms in
the interaction with the target, say the $Q( {\bm x}_1, {\bm y}_1 ,
{\bm x}_2 , {\bm y}_2 )$ in \eq{crossedinteraction}.  Among many terms
which contribute to the quadrupole amplitude in \eq{eq:quad_exp},
there is a term
\begin{align}
  \label{eq:quad}
  Q( {\bm x}_1, {\bm y}_1 , {\bm x}_2 , {\bm y}_2 )
  \bigg|_{\mbox{order}-Q_{s0}^4} \sim ({\bm x}_1 - {\bm y}_1)^2 \,
  Q_{s0}^2 \, \ln \left( \frac{1}{|{\bm x}_1 - {\bm y}_1| \, \Lambda}
  \right) \ ({\bm x}_2 - {\bm y}_2)^2 \, Q_{s0}^2 \, \ln \left(
    \frac{1}{|{\bm x}_2 - {\bm y}_2| \, \Lambda} \right) + \ldots \, .
\end{align}
Using \eq{eq:quad} in \eq{crossed_xsect} and changing the interaction
variables to those defined in \eq{eq:change} we see that the remaining
${\bm b}_1$ and ${\bm b}_2$ integrals, after shifting those variables
by $\bm B$, become simple Fourier transforms of the projectile nucleus
profile function $T_1 ({\bm b})$,
\begin{align}
  \label{eq:fourier}
  \int d^2 b_1 \, d^2 b_2 \, T_1 ({\bm b}_1) \, T_1 ({\bm b}_2) \,
  e^{- i \, ({\bm k}_1 - {\bm k}_2) \cdot ({\bm b}_1-{\bm b}_2)} +
  ({\bm k}_2 \rightarrow - {\bm k}_2).
\end{align}
In the limit of a sufficiently large projectile nucleus the Fourier
transforms give a delta-function, yielding
\begin{align}
  \label{eq:hbt1}
  \frac{d \sigma_{HBT}}{d^2 k_1 dy_1 d^2 k_2 dy_2} \sim \delta^2 ({\bm
    k}_1 - {\bm k}_2) + \delta^2 ({\bm k}_1 + {\bm k}_2),
\end{align}
in agreement with the result in \eq{eq:HBT_LO}. Our present estimate
shows that a more careful evaluation of \eq{eq:fourier} would give a
smother peak in $|{\bm k}_1 - {\bm k}_2|$ at ${\bm k}_1 = {\bm k}_2$
with the width determined by the inverse radius of the projectile
nucleus $R_1$.

\begin{figure}[th]
  \includegraphics[width= 0.45 \textwidth]{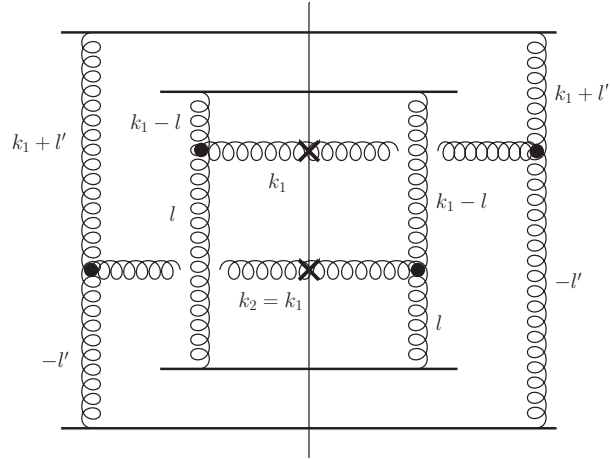}
  \caption{An example of a lowest-order diagram generating HBT-type
    correlations.}
\label{hbt_graph} 
\end{figure}

The first term on the right-hand-side of \eq{eq:hbt1} (or
\eq{eq:fourier}) has the trademark form of the Hanbury Brown--Twiss
(HBT) correlations \cite{HanburyBrown:1956pf}, which are widely
studied in heavy ion physics
\cite{Heinz:1999rw,Adams:2003ra,Kopylov:1972qw}. These correlations
are normally local both in rapidity ($y_1 = y_2$) and in the
transverse momentum ${\bm k}_1 = {\bm k}_2$: however, in
\eq{eq:HBT_LO} (or, in \eq{eq:hbt1}) we only have locality in
transverse momenta. This can be explained by the fact that the extent
of the interaction region in the longitudinal direction, commonly
labeled $R_{long}$, is very small in our case due to the extreme
Lorentz-contraction of the two colliding nuclei leading to smearing of
the longitudinal HBT correlations.

We conclude that the correlations resulting from \eq{crossed_xsect}
include HBT. Diagrammatic representation of the HBT correlations was
discussed earlier in \cite{Capella:1991mp}. An example of the diagram
giving rise to the correlations in \eq{eq:HBT_LO} is shown in
\fig{hbt_graph}.

Interestingly though, due to the ${\bm k}_2 \leftrightarrow - {\bm
  k}_2$ symmetry of the two-gluon production cross section
\eqref{crossed_xsect}, the HBT peak at ${\bm k}_1 = {\bm k}_2$ in
\eq{eq:HBT_LO} is accompanied by an identical peak at ${\bm k}_1 = -
{\bm k}_2$ resulting from $\delta^2 ({\bm k}_1 + {\bm k}_2)$. We thus
obtain a back-to-back HBT correlation resulting from multiple
rescatterings in nuclei. (The origin of our back-to-back HBT
correlations is different from that of the back-to-back HBT-like
correlations proposed in \cite{Asakawa:1998cx,Csorgo:2007iv}.)

Note that it is possible that the process of hadronization would
affect the phases of the produced gluons, possibly destroying the
perturbative HBT correlations of \eq{crossed_xsect} and replacing them
with the HBT correlations of the non-perturbative origin. The same
hadronization process may also destroy the back-to-back HBT
correlations from \eq{eq:HBT_LO}.


\section{Conclusion}
\label{sec:conc}

Above we have derived the cross section for two-gluon production in
the heavy-light ion collisions. The cross section is given by
\eq{eq_all}, with the two contributions shown in
Eqs.~\eqref{eq:2glue_prod_main} and
\eqref{crossed_xsect}. Concentrating on gluon production with a
long-range separation in rapidity of the two gluons we evaluated the
interaction with the target in \eq{eq:2glue_prod_main} using the MV/GM
approximation in \eq{Delta_exp}. The interaction with the target in
\eq{crossed_xsect} in the MV-model framework can be obtained from
Eqs.~\eqref{eq:SG_GM} and \eqref{eq:quad_MV}.

\begin{figure}[h]
  \includegraphics[width= 0.45 \textwidth]{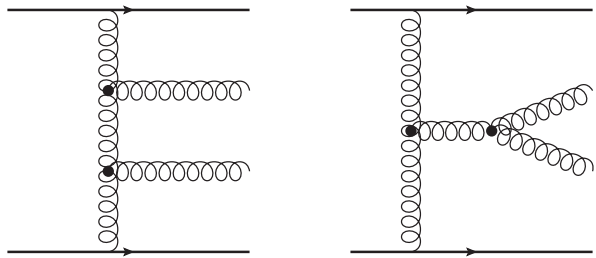}
  \caption{Two-gluon production in a standard pQCD formalism, with the
    diagram on the left generating back-to-back correlations, and the
    diagram on the right responsible for the near-side collinear
    correlations.}
  \label{pQCD}
\end{figure}

Analyzing the corresponding correlation function we identified four
main types of correlations: (i) geometric correlations, (ii) HBT
correlations, (iii) away-side correlations, and (iv) the near-side
correlations. HBT correlations (ii) are local in rapidity and thus are
outside the main scope of our work here: however, we find that HBT
correlations arising in our calculation are accompanied by a
back-to-back HBT peak, as follows from \eq{eq:HBT_LO}. Geometric
correlations (i) arise in \eq{corr_fact} due to nucleons being
confined within the same nucleus: more work is needed to see how
strong they are and what role they are playing in experimental data.

Concentrating on the long-range rapidity correlations (iii) and (iv)
we have shown that the two correlations have identical azimuthal
profiles centered around $\Delta \phi = \pi$ and $\Delta \phi = 0$
correspondingly. This result is true to all orders of the calculation
in the heavy-light ion collision considered here. Let us stress here
the difference between this result and the regular jet correlations:
indeed a two-hadron production cross section calculated in
perturbative QCD (pQCD), illustrated by the examples in \fig{pQCD},
contains both the back-to-back and near-side correlations. However,
while the jet back-to-back correlations are long-range in rapidity,
the jet near-side correlations are local in rapidity, being due to
collinear jet showers around the trigger hadron (see
\cite{Kovchegov:2002nf,Kovchegov:2002cd} for an analysis of mini-jet
correlations in heavy ion collisions). Hence in standard pQCD
formalism the strengths along with rapidity and azimuthal shapes of
the back-to-back and near-side correlations are different. This is in
stark contrast to our correlations (iii) and (iv), which are both
long-range in rapidity, and have identical amplitudes and
rapidity/azimuthal shapes. The correlations (iii) and (iv) arise due
to multiple rescatterings in both nuclei and are, therefore, typical
of the ion--ion collisions, weakening in the limit when one (or both)
of the ions is (are) small (a proton). It is possible that this
correlation would complicate the experimental extraction of the
elliptic flow observable $v_2$ due to the QGP flow in heavy ion
collisions, though it may also be that the correlation is too weak to
present a significant background. More detailed numerical analysis of
our result is needed to clarify this point.


\section*{Acknowledgments}

The authors are grateful to Adrian Dumitru, Ulrich Heinz, Jamal
Jalilian-Marian, Mike Lisa, and Kirill Tuchin for inspiring
discussions of two-particle correlations.

This research is sponsored in part by the U.S. Department of Energy
under Grant No. DE-SC0004286. \\

After this paper had been posted online as a preprint, a new
experimental result \cite{Abelev:2012aa} appeared, reporting on the
discovery of identical near- and away-side correlations found in the
$p+Pb$ collisions at the LHC, in qualitative agreement with one of the
main results of this work. \\


\section{Appendix}
\renewcommand{\theequation}{A\arabic{equation}}
  \setcounter{equation}{0}

Our goal here is to calculate the following object
\begin{align}
\label{corr1}
  \frac{1}{(N_c^2 -1)^2} \, \langle Tr[ U_{{\bm x}_1} U_{{\bm
      x}_2}^\dagger ] \, Tr[ U_{{\bm x}_3} U_{{\bm x}_4}^\dagger ]
  \rangle
\end{align}
in the quasi-classical MV/GM approximation at the lowest non-trivial
order in $1/N_c^2$ expansion in the 't Hooft's large-$N_c$ limit.  We
assume that the saturation scale is $N_c$-independent, that is
$Q_{s0}^2 \sim (N_c)^0$, due to an order-$N_c^2$ number of ``valence''
partons in each nucleon in the target nucleus.

\begin{figure}[h]
  \includegraphics[width= 0.45 \textwidth]{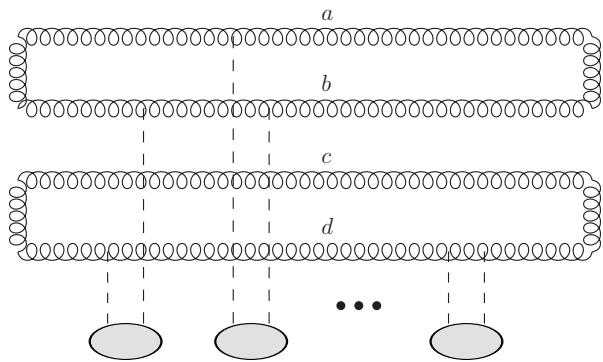}
  \caption{Diagrams contributing to the correlator in \eq{corr1} in
    the quasi-classical approximation. Vertical dashed lines denote
    $t$-channel gluons, while the shaded ovals represent the nucleons
    in the target nucleus.}
\label{trtr} 
\end{figure}

A sample of the diagrams contributing to \eq{corr1} is shown in
\fig{trtr}, where the $t$-channel gluons are denoted by dashed lines
to simplify the picture. The calculation is straightforward
\cite{Kovchegov:2001ni}: one simply has to exponentiate the two-gluon
exchange interaction with a single nucleon (the nucleons are denoted
by shaded ovals in \fig{trtr}). The only complication is that, unlike
the dipole amplitude calculated in \cite{Mueller:1989st}, the
interaction now, for the double-trace operator \eqref{corr1}, is a
matrix in the color space. A double trace operator like \eqref{corr1}
with the Wilson lines in the fundamental representation was calculated
earlier in \cite{Marquet:2010cf} (for similar calculations see also
\cite{Nikolaev:2005zj,Dominguez:2011gc,Iancu:2011ns}).

To exponentiate the matrix we have to choose a basis in the color
space of four $s$-channel gluons in \fig{trtr}: clearly the net color
of the four gluons is always zero. The color states of the four gluons
can be classified according to the color states of the top two
$s$-channel gluons, since the color of the bottom pair of $s$-channel
gluons is determined by requiring net color-neutrality of the four
$s$-channel gluon system. The colors of a gluon pair can be decomposed
in the following irreducible representations
\begin{align}
  \label{8x8}
  & {(N_c^2 - 1)} \otimes {(N_c^2 - 1)} = V_1 \oplus V_2 \oplus V_3
  \oplus V_4 \oplus V_5 \oplus V_6 \oplus V_7 \notag \\ & = {\bm 1}
  \oplus {(N_c^2 -1)} \oplus \frac{N_c^2 (N_c -3) (N_c +1)}{4} \oplus
  \frac{N_c^2 (N_c + 3) (N_c - 1)}{4} \oplus {(N_c^2 -1)} \oplus
  \frac{(N_c^2 -1) (N_c^2 - 4)}{4} \oplus \frac{(N_c^2 -1) (N_c^2 -
    4)}{4}.
\end{align}
Here we are following the notations introduced in
\cite{Cvitanovic:2008zz}, see page 120 there. Labeling the colors of
the four $s$-channel gluons in an arbitrary color state by $a, b, c$,
and $d$ as shown in \fig{trtr} we define the color states
corresponding to representations $V_1$, $V_2$, and $V_5$ by
\begin{align}
  \label{eq:color_states}
  |P_1 \rangle = \frac{1}{N_c^2 -1} \, \delta^{ab} \, \delta^{cd} , \
  \ \ |P_2 \rangle = \frac{1}{\sqrt{N_c^2 -1}} \, \frac{N_c}{N_c^2 -4}
  \, d^{abe} \, d^{cde}, \ \ \ |P_5 \rangle = \frac{1}{\sqrt{N_c^2
      -1}} \, \frac{1}{N_c} \, f^{abe} \, f^{cde},
\end{align}
where we differ from $P_i$'s in \cite{Cvitanovic:2008zz} by
prefactors, since here we demanded that our color states are
normalized to one, $\langle P_i | P_j \rangle = \delta^{ij} $. Other
color states can be constructed as well \cite{Cvitanovic:2008zz}, but
we will only need the states in \eq{eq:color_states} for the
calculation below.

We denote by $\hat M$ the interaction with a single nucleon by a
two-gluon exchange: it is a matrix in the color space of the
$s$-channel gluons. Since in the correlator \eqref{corr1} the top
(bottom) two $s$-channel gluons are in the color-singlet state both
before and after the interaction, we write
\begin{align}
  \label{corr2}
  \frac{1}{(N_c^2 -1)^2} \, \langle Tr[ U_{{\bm x}_1} U_{{\bm
      x}_2}^\dagger ] \, Tr[ U_{{\bm x}_3} U_{{\bm x}_4}^\dagger ]
  \rangle = \langle P_1 | e^{\hat M} | P_1 \rangle. 
\end{align}
Expanding the exponential in a power series and inserting unit
operators ${\bm 1} = \sum_i |P_i \rangle \, \langle P_i |$ between all
the $\hat M$'s yields
\begin{align}
  \label{corr3}
  \frac{1}{(N_c^2 -1)^2} \, \langle Tr[ U_{{\bm x}_1} U_{{\bm
      x}_2}^\dagger ] \, Tr[ U_{{\bm x}_3} U_{{\bm x}_4}^\dagger ]
  \rangle = \left( e^M \right)_{11}
\end{align}
where the $7 \times 7$ matrix $M$ is defined by its elements,
\begin{align}
  \label{Mdef}
  M_{ij} = \langle P_i | {\hat M} | P_j \rangle .  
\end{align}
All one has to do now is to find the matrix $M$ from \eq{Mdef},
exponentiate it, picking up the ``$11$'' matrix element of the
exponential. Such a calculation, while straightforward, is rather
involved: here we will utilize the large-$N_c$ limit to construct an
approximate result.

Calculating some of the elements of the matrix $M$ and evaluating the
$N_c$-order of the remaining matrix elements yields
\begin{align}
  \label{eq:Matrix}
  M = \left( \begin{array}{ccccccc}
      D_1 & 0 & 0 & 0 & \frac{D_3-D_2}{\sqrt{N_c^2-1}} & 0 & 0 \\
      0 & \frac{1}{2}D_1 + \frac{1}{4}(D_2+D_3) & O(1/N_c) & O (1/N_c) & \frac{1}{4}(D_3-D_2) & O (1/N_c) & O (1/N_c) \\
      0 & O (1/N_c) & \ldots & \ldots & O (1/N_c) & \ldots & \ldots \\
      0 & O (1/N_c) & \ldots & \ldots & O (1/N_c) & \ldots & \ldots \\
      \frac{D_3-D_2}{\sqrt{N_c^2-1}} & \frac{1}{4}(D_3-D_2) & O
      (1/N_c) & O (1/N_c) & \frac{1}{2}D_1 + \frac{1}{4}(D_2+D_3)& O
      (1/N_c) & O (1/N_c) \\
     0 & O (1/N_c) & \ldots & \ldots & O (1/N_c) & \ldots & \ldots \\
      0 & O (1/N_c) & \ldots & \ldots & O (1/N_c) & \ldots & \ldots
\end{array} \right)
\end{align}
with $D_i$ defined in Eqs.~(\ref{Ds}) and ellipsis denoting the matrix
elements we do not need to calculate as they are at most order-1 in
$N_c$ counting. From \eq{eq:Matrix} we see that if we start in the
color-singlet state $|P_1 \rangle$ for the top (bottom) gluon pair, a
single interaction can either leave the two gluons in the
color-singlet state, or can flip them into a color-octet state $|P_5
\rangle$. The latter transition comes in with an order-$1/N_c$
suppression factor. In order to evaluate \eqref{corr1} we have to
start and finish with a color-singlet state: to order-$1/N_c^2$ we may
have at most two such transitions: $|P_1 \rangle \rightarrow |P_5
\rangle$ and the inverse, $|P_5 \rangle \rightarrow |P_1
\rangle$. Once the system is in the color-octet $|P_5 \rangle$ state,
it can continue its random walk through color space: however, if we
want to keep the calculation at the order-$1/N_c^2$, the interactions
between the two transitions should be leading-order in $N_c$. This is
why only the leading-$N_c$-order matrix elements $M_{22}$, $M_{25} =
M_{52}$, and $M_{55}$ contribute in the $6 \times 6$ matrix $M_{ij}$
with $i,j = 2, \ldots, 7$: we do not need to calculate the
$1/N_c$-suppressed elements in \eq{eq:Matrix} or the elements denoted
by ellipsis which can not contribute.

Exponentiating the matrix $M$ from \eq{eq:Matrix}, picking up the
``$11$'' element of the obtained matrix and expanding the result to
order-$1/N_c^2$ yields
\begin{align}
  \label{eq:ans}
  \frac{1}{(N_c^2 -1)^2} \, \langle Tr[ U_{{\bm x}_1} U_{{\bm
      x}_2}^\dagger ] \, Tr[ U_{{\bm x}_3} U_{{\bm x}_4}^\dagger ]
  \rangle = e^{D_1} + \frac{(D_3-D_2)^2}{N_c^2} \, & \left[
    \frac{e^{D_1}}{D_1-D_2} - \frac{2 \, e^{D_1}}{(D_1-D_2)^2} +
    \frac{e^{D_1}}{D_1-D_3} - \frac{2 \,
      e^{D_1}}{(D_1-D_3)^2} \right. \notag \\
  & \left. + \frac{2 \, e^{\frac{1}{2}(D_1+D_2)}}{(D_1-D_2)^2} +
    \frac{2 \, e^{\frac{1}{2}(D_1+D_3)}}{(D_1-D_3)^2} \right] + O
  \left( \frac{1}{N_c^4} \right).
\end{align}
Finally, noting that (see \eq{eq:SG_GM})
\begin{align}
  \label{eq:factDel}
  \frac{1}{(N_c^2 -1)^2} \, \langle Tr[ U_{{\bm x}_1} U_{{\bm
      x}_2}^\dagger ] \rangle \, \langle Tr[ U_{{\bm x}_3} U_{{\bm
      x}_4}^\dagger ] \rangle =  e^{D_1}
\end{align}
and using \eq{Ddef} we obtain \eq{Delta_exp}.



\providecommand{\href}[2]{#2}\begingroup\raggedright\endgroup


\end{document}